# Bias-Corrected and Variance-Corrected MLE for the New Median Based Unit Weibull Distribution (MBUW)

Iman M. Attia *

[Imanattiathesis1972@gmail.com](mailto:Imanattiathesis1972@gmail.com) ,[imanattia1972@gmail.com](mailto:imanattia1972@gmail.com)

*Department of Mathematical Statistics, Faculty of Graduate Studies for Statistical Research, Cairo University, Egypt*

***Abstract***: As the maximum likelihood method is the most commonly used method for parameter estimation being unbiased, consistent, efficient, and asymptotically normal, MLE is used to fit the new distribution (MBUW). But in small to moderate sample sizes, this MLE estimator is biased unlike the MLE estimators obtained from large sample sizes. In this paper, the Bias-corrected approach for this distribution is discussed and applied to real data analysis. The MLE estimators of MBUW obtained from some optimization techniques like the derivative free Nelder Mead algorithm suffer from a significantly high correlation that is reflected in high covariance between the parameters. Also, this association between the parameters affects the variances which may be inflated enough to approach infinity hampering the construction of confidence intervals for each parameter. This problem may arise with any optimization technique which necessitates remedies trying to fix it. The author also elaborates on a variance correction approach heavily relying on re-parameterizing the negative log-likelihood.

# Introduction

The most commonly used parameter estimation method is the maximum likelihood estimation (MLE), which is crucial for any probability distribution (Pawitan, 2001),(Millar, 2011). MLE estimator has many advantageous properties such as asymptotically unbiased, consistent, efficient and asymptotically normally distributed. These properties depend mainly on large sample size. In small to moderate sample size some of these properties like un-biasness may be violated. The error obtained from the difference between the expected value of the parameter and the true value of the parameter is reduced as the sample size increases. For that reason, the researchers develop almost unbiased estimators for different distributions. To mention some of them: (Saha & Paul, 2005),(Cordeiro et al., 1997), (Giles, 2012) (Cribari-Neto & Vasconcellos, 2002), (Lemonte et al., 2007), (Giles, &

Feng, 2009),(Giles, 2012), (Schwartz et al., 2013), (Giles et al., 2013), (Teimouri & Nadarajah, 2013), (Zhang & Liu, 2017), (Singh et al., 2015), (Lagos-Àlvarez, et al., 2011), (Schwartz & Giles, 2016), (Wang & Wang, 2017), (Ling & Giles, 2014), (Lemonte, 2011), (Mazucheli & Dey, 2018), (Reath, 2016), and (Teimouri & Nadarajah, 2016), and references cited therein.

It is possible to approximate the bias of the MLE of the estimated parameter for a single parameter distribution to the $O(n^{-1})$ even if the estimated parameter is not in a closed form expression (Bartlett, 1953a) (Haldane & Smith, 1956), (Bartlett, 1953b) and (Haldane, 1953)derived the analytic approximations for two-parameters log-likelihood functions utilizing the Tayler series expansions that may be tough for multi-parameter distributions as illustrated by (Shenton & Bowman, 1963).

Many approaches have been proposed to correct the bias for MLE. The first approach is called the "corrective approach". It is the analytical methodology advocated by (Cox & Snell, 1968). It is an analytical expression for the bias to $O(n^{-1})$ of the MLE estimators, then using these expressions to bias-correct the MLE estimator yielding estimators that are unbiased to $O(n^{-2})$. The second approach is parametric Bootstrap resampling technique advised by (Efron, 1982). It is also a second-order bias correction. The bias-correction is carried out numerically without developing analytical expression for the bias function. The third approach is called "preventive approach" recommended by (Firth, 1993). It is analytic procedure that mandates modification of the score function of the log-likelihood function before solving for the MLEs and it reduces the bias to the order $O(n^{-2})$. The first two approaches have unsophisticated mathematical expressions which renders them appealing and simple to calculate.

The BMUW distribution has been discussed in earlier work by the author (Iman M. Attia, 2024) as regards properties and some methods of estimation with applications on real data analysis. The new distribution has the following PDF, CDF and quantile function respectively, as shown in equations (1-3)

$$f(y) = \frac{6}{\alpha^\beta} \left[1 - y^{\frac{1}{\alpha^\beta}}\right] y^{\left(\frac{2}{\alpha^\beta}-1\right)} , \quad 0 < y < 1, \quad \alpha > 0, \beta > 0 \ldots\ldots\ldots\ldots\ldots\ldots\ldots\ldots\ldots\ldots (1)$$

$$F(y) = 3y^{\frac{2}{\alpha^\beta}} - 2y^{\frac{3}{\alpha^\beta}} , \quad 0 < y < 1, \quad \alpha > 0, \beta > 0 \ldots\ldots\ldots\ldots\ldots\ldots\ldots\ldots\ldots\ldots (2)$$

$$u = F(y) = 3y^{\frac{2}{\alpha^\beta}} - 2y^{\frac{3}{\alpha^\beta}} = -2\left(y^{\frac{1}{\alpha^\beta}}\right)^3 + 3\left(y^{\frac{1}{\alpha^\beta}}\right)^2 \ldots\ldots\ldots\ldots\ldots\ldots\ldots\ldots (3)$$

This distribution is defined on the unit interval. It can accommodate skew data expressed as proportions. The distribution can have a bathtub or unimodal shapes according to the parameters.

In this paper, the author discusses the bias-corrected MLE approach and the variance-corrected MLE technique. The paper is constructed into sections. Methods are contained in section1 and section 2. In section 1, the author explains the corrective procedure for the bias of the MLE estimators. In section 2, the author derives the analytic function for the bias of MLE estimators for the MBUW distribution. Results are revealed in section 3. In section 3, the author describes the variance-corrected MLE procedure for MBUR and implements this approach on real data. Discussion is carried out in section 4. In section 4, the author deploys the bias-corrected MLE on real data comparing values of the estimated parameters obtained from the variance-corrected MLE with the values obtained by bias-corrected approach. In section 5, conclusions and recommendations are elucidated.

# Methods

## Section 1

## Bias-corrected MLE

Let $\Theta$ be a p-dimensional unknown parameter vector and $l = l(\Theta|y)$ be the log-likelihood function for a sample of n observations. Assume this log-likelihood is regular with respect to all derivatives up to and including those of third order. The joint cumulants of the log-likelihood derivatives are defined as follows in equations (4-6)

$$k_{ij} = E\left(\frac{\partial^2 l}{\partial \Theta_i \partial \Theta_j}\right) ; i,j = 1,2,\ldots,p \quad \ldots\ldots\ldots\ldots\ldots\ldots\ldots\ldots\ldots\ldots\ldots\ldots\ldots (4)$$

$$k_{ijl} = E\left(\frac{\partial^3 l}{\partial \Theta_i \partial \Theta_j \partial \Theta_l}\right); i,j = 1,2,\ldots,p \quad \ldots\ldots\ldots\ldots\ldots\ldots\ldots\ldots\ldots\ldots\ldots\ldots (5)$$

$$k_{ij,l} = E\left(\frac{\partial^2 l}{\partial \Theta_i \partial \Theta_j} \frac{\partial^3 l}{\partial \Theta_l}\right) ; i,j = 1,2,\ldots,p \quad \ldots\ldots\ldots\ldots\ldots\ldots\ldots\ldots\ldots\ldots\ldots (6)$$

The derivative of these cumulants are defined as in equation (7)

$$k_{ij}^{(l)} = E\left(\frac{\partial k_{ij}}{\partial \Theta_l}\right); i,j,l = 1,2,\ldots,p \quad \ldots\ldots\ldots\ldots\ldots\ldots\ldots\ldots\ldots\ldots\ldots\ldots (7)$$

The expressions in equations (4-7) are assumed to be $O(n)$.

(Cox & Snell, 1968) revealed that when the sample data are independent (but not necessarily identically distributed) the bias of the $s^{th}$ element of the MLE of $\widehat{\Theta}_s$, is calculated as

$$Bias(\hat{\Theta}_s) = \sum_{i=1}^{p}\sum_{j=1}^{p}\sum_{l=1}^{p} k^{si}k^{jl}\left[0.5 * k_{ijl} + k_{ij,l}\right] + O(n^{-2}) \ldots \ldots \ldots \ldots \ldots \ldots \ldots \ldots (8)$$

where: $s = 1, \ldots, p$ and $k^{ij}$ is the $(i,j)th$ element of the inverse of the expected information matrix $K = -k_{ij}$.

(Cordeiro & Klein, 1994) consequently established that equation (8) yet maintains if the data are non-independent and that it can be disclosed as

$$Bias(\hat{\Theta}_s) = \sum_{i=1}^{p} k^{si} \sum_{j=1}^{p}\sum_{l=1}^{p}\left[k_{ij}^{(l)} - 0.5 * k_{ijl}\right] k^{jl} + O(n^{-2}) \ldots \ldots \ldots \ldots \ldots \ldots \ldots (9)$$

The bias equation in (9) is largely simpler to calculate than in (8), because it does not include terms of the form given in (6).

Defining the following terms in equations (10-11):

$$a_{ij}^{(l)} = k_{ij}^{(l)} - 0.5 k_{ijl} \; ; \quad i,j,l = 1,2,\ldots,p \ldots \ldots \ldots \ldots \ldots \ldots \ldots \ldots \ldots \ldots (10)$$

$$A^{(l)} = \left\{a_{ij}^{(l)}\right\} \; ; \quad i,j,l = 1,2,\ldots,p \ldots \ldots \ldots \ldots \ldots \ldots \ldots \ldots \ldots \ldots (11)$$

and collecting terms up into matrices $A = \left[\; A^{(1)} | \ldots | A^{(p)} \;\right]$

The $O(n^{-2})$ bias of the MLE of $\hat{\Theta}$ in (9) can be rephrased in the convenient form equation (12):

$$Bias(\hat{\Theta}_s) = \hat{K}^{-1}\hat{A}\, vec\left(\hat{K}^{-1}\right) \ldots \ldots \ldots \ldots \ldots \ldots \ldots \ldots \ldots \ldots \ldots \ldots (12)$$

where $\hat{K} = K|_{\Theta=\hat{\Theta}}$ and $\hat{A} = A|_{\Theta=\hat{\Theta}}$, the value of $\hat{\Theta}_s$ is obtained by solving the roots of log-likelihood equations using the numerical methods. $Vec(.)$ means vectorization operator, which stacks the columns of the matrix in question one above the other, forming one extended column vector. Hence the bias adjusted-MLE is defined in equation (13) as

$$\tilde{\Theta} = \hat{\Theta} - \hat{K}^{-1}\hat{A}\, vec\left(\hat{K}^{-1}\right) \ldots \ldots \ldots \ldots \ldots \ldots \ldots \ldots \ldots \ldots \ldots \ldots (13)$$

One of the benefits of this method is that these expressions can be calculated when solving for the root of the log-likelihood equations do not disclose an analytic closed-form solution. In such circumstances, the bias-corrected MLE can be easily obtained by means of standard numerical methods, and $\tilde{\Theta}$ is unbiased $O(n^{-2})$

# Section 2

# Bias Reduction and the MBUW

The PDF of MBUW satisfies the regularity conditions. The first order partial derivatives of log-likelihood with respect to alpha and beta parameters are defined in equation (14-15) as follows:

$$\frac{\partial l}{\partial \alpha} = \frac{\beta}{\alpha} + \sum_{i=1}^{n} \frac{y^{\alpha^{-\beta}} \beta (\ln y) \alpha^{(-\beta-1)}}{1 - y^{\alpha^{-\beta}}} - 2\beta \alpha^{(-\beta-1)} \sum_{i=1}^{n} \ln y \dots \dots \dots (14)$$

$$\frac{\partial l}{\partial \beta} = -n \ln \alpha + \sum_{i=1}^{n} \frac{y^{\alpha^{-\beta}} \alpha^{-\beta} (\ln y)(\ln \alpha)}{1 - y^{\alpha^{-\beta}}} - 2\alpha^{-\beta} (\ln \alpha) \sum_{i=1}^{n} \ln y \dots \dots \dots (15)$$

Equations (16-24) are the higher order derivatives (for one observation):

$$\frac{\partial^2 l}{\partial \alpha^2} = \frac{\beta}{\alpha^2} - \left( \frac{y^{\alpha^{-\beta}} \beta^2 (\ln y)^2 \alpha^{(-2\beta-2)}}{1 - y^{\alpha^{-\beta}}} \right) - \frac{y^{\alpha^{-\beta}} \beta(\beta+1)(\ln y)\alpha^{(-\beta-2)}}{1 - y^{\alpha^{-\beta}}}$$

$$- \left( \frac{y^{2\alpha^{-\beta}} \beta^2 (\ln y)^2 \alpha^{(-2\beta-2)}}{\left(1 - y^{\alpha^{-\beta}}\right)^2} \right) + 2\beta(\beta+1)\alpha^{(-\beta-2)}(\ln y) \dots \dots \dots (16)$$

$$\frac{\partial^2 l}{\partial \beta^2} = \left( \frac{-y^{\alpha^{-\beta}} (\ln \alpha)^2 (\ln y)^2 \alpha^{(-2\beta)}}{1 - y^{\alpha^{-\beta}}} \right) - \left( \frac{y^{\alpha^{-\beta}} (\ln \alpha)^2 (\ln y)\alpha^{(-\beta)}}{1 - y^{\alpha^{-\beta}}} \right)$$

$$- \left( \frac{y^{2\alpha^{-\beta}} (\ln \alpha)^2 (\ln y)^2 \alpha^{(-2\beta)}}{\left(1 - y^{\alpha^{-\beta}}\right)^2} \right) + 2\alpha^{(-\beta)}(\ln y)(\ln \alpha)^2 \dots \dots \dots (17)$$

$$\frac{\partial^2 l}{\partial \alpha \partial \beta} = \frac{-1}{\alpha} - \left( \frac{y^{\alpha^{-\beta}} \beta (\ln \alpha)(\ln y)^2 \alpha^{(-2\beta-1)}}{1 - y^{\alpha^{-\beta}}} \right) + \frac{y^{\alpha^{-\beta}} (\ln y)\alpha^{(-\beta-1)}}{1 - y^{\alpha^{-\beta}}}$$

$$- \left( \frac{y^{\alpha^{-\beta}} \beta (\ln \alpha)(\ln y)\alpha^{(-\beta-1)}}{1 - y^{\alpha^{-\beta}}} \right) - \left( \frac{y^{2\alpha^{-\beta}} \beta (\ln \alpha)(\ln y)^2 \alpha^{(-2\beta-1)}}{\left(1 - y^{\alpha^{-\beta}}\right)^2} \right)$$

$$-2\alpha^{(-\beta-1)} (\ln y) + 2\beta \alpha^{(-\beta-1)} (\ln \alpha)(\ln y) \dots \dots \dots (18)$$

$$\frac{\partial^3 l}{\partial \alpha^2 \partial \beta} = \frac{1}{\alpha^2} + \frac{y^{\alpha-\beta}\beta^2(\ln y)^3 (\ln \alpha)\, \alpha^{(-3\beta-2)}}{1-y^{\alpha-\beta}} - \frac{2\, y^{\alpha-\beta}\beta(\ln y)^2 \alpha^{(-2\beta-2)}}{1-y^{\alpha-\beta}}$$

$$+ \frac{2y^{\alpha-\beta}\beta^2(\ln y)^2 (\ln \alpha)\, \alpha^{(-2\beta-2)}}{1-y^{\alpha-\beta}} + \frac{y^{2\alpha-\beta}\beta^2(\ln y)^3 (\ln \alpha)\, \alpha^{(-3\beta-2)}}{\left(1-y^{\alpha-\beta}\right)^2}$$

$$- \frac{y^{\alpha-\beta}(2\beta+1)(\ln y)\, \alpha^{(-\beta-2)}}{1-y^{\alpha-\beta}} + \frac{y^{\alpha-\beta}(\beta^2+\beta)(\ln \alpha)(\ln y)^2 \alpha^{(-2\beta-2)}}{1-y^{\alpha-\beta}}$$

$$+ \frac{y^{\alpha-\beta}(\beta^2+\beta)(\ln \alpha)(\ln y)\alpha^{(-\beta-2)}}{1-y^{\alpha-\beta}} + \frac{y^{2\alpha-\beta}(\beta^2+\beta)(\ln y)^2 (\ln \alpha)\, \alpha^{(-2\beta-2)}}{\left(1-y^{\alpha-\beta}\right)^2}$$

$$+ \frac{2\, y^{2\alpha-\beta}\beta^2(\ln y)^3 (\ln \alpha)\, \alpha^{(-3\beta-2)}}{\left(1-y^{\alpha-\beta}\right)^2} - \frac{2y^{2\alpha-\beta}\beta(\ln y)^2\, \alpha^{(-2\beta-2)}}{\left(1-y^{\alpha-\beta}\right)^2}$$

$$+ \frac{2\, y^{2\alpha-\beta}\beta^2(\ln y)^2 (\ln \alpha)\, \alpha^{(-2\beta-2)}}{\left(1-y^{\alpha-\beta}\right)^2} + \frac{2y^{3\alpha-\beta}\beta^2 (\ln \alpha)(\ln y)^3\, \alpha^{(-3\beta-2)}}{\left(1-y^{\alpha-\beta}\right)^3}$$

$$+ 4\beta\alpha^{(-\beta-2)}(\ln y) + 2\alpha^{(-\beta-2)}(\ln y) - 2\beta^2\alpha^{(-\beta-2)}(\ln y)(\ln \alpha)$$

$$-2\beta\alpha^{(-\beta-2)}(\ln y)(\ln \alpha) \quad \ldots \ldots \ldots \ldots \ldots \ldots \ldots \ldots \ldots \ldots \ldots \ldots \ldots \ldots \ldots (19)$$

$$\frac{\partial^3 l}{\partial \beta^2 \partial \alpha} = \frac{y^{\alpha-\beta}\beta(\ln y)^3 (\ln \alpha)^2 \alpha^{(-3\beta-1)}}{1-y^{\alpha-\beta}} + \frac{2\, y^{\alpha-\beta}\beta(\ln y)^2 (\ln \alpha)^2 \alpha^{(-2\beta-1)}}{1-y^{\alpha-\beta}}$$

$$- \frac{2y^{\alpha-\beta}(\ln y)^2 (\ln \alpha)\, \alpha^{(-2\beta-1)}}{1-y^{\alpha-\beta}} + \frac{y^{2\alpha-\beta}\beta(\ln y)^3 (\ln \alpha)^2 \alpha^{(-3\beta-1)}}{\left(1-y^{\alpha-\beta}\right)^2}$$

$$+ \frac{y^{\alpha-\beta}\beta\,(\ln y)^2 (\ln \alpha)^2\, \alpha^{(-2\beta-1)}}{1-y^{\alpha-\beta}} + \frac{y^{\alpha-\beta}(\beta)(\ln y)(\ln \alpha)^2 \alpha^{(-\beta-1)}}{1-y^{\alpha-\beta}}$$

$$- \frac{2y^{\alpha-\beta}\,(\ln \alpha)(\ln y)\alpha^{(-\beta-1)}}{1-y^{\alpha-\beta}} + \frac{y^{2\alpha-\beta}(\beta)(\ln y)^2 (\ln \alpha)^2 \alpha^{(-2\beta-1)}}{\left(1-y^{\alpha-\beta}\right)^2}$$

$$+ \frac{2\, y^{2\alpha-\beta}\beta(\ln y)^3 (\ln \alpha)^2 \alpha^{(-3\beta-1)}}{\left(1-y^{\alpha-\beta}\right)^2} + \frac{2y^{2\alpha-\beta}\beta(\ln y)^2\,(\ln \alpha)^2\alpha^{(-2\beta-1)}}{\left(1-y^{\alpha-\beta}\right)^2}$$

$$- \frac{2\, y^{2\alpha-\beta}(\ln y)^2 (\ln \alpha)\, \alpha^{(-2\beta-1)}}{\left(1-y^{\alpha-\beta}\right)^2} + \frac{2y^{3\alpha-\beta}\beta\,(\ln \alpha)^2 (\ln y)^3\, \alpha^{(-3\beta-1)}}{\left(1-y^{\alpha-\beta}\right)^3}$$

$$-2\beta\alpha^{(-\beta-1)}(\ln y)(\ln \alpha)^2 + 4\alpha^{(-\beta-1)}(\ln y)(\ln \alpha) \quad \ldots \ldots \ldots \ldots \ldots \ldots \ldots \ldots \ldots \ldots \ldots (20)$$

$$\frac{\partial^3 l}{\partial \alpha^3} = \frac{-2\beta}{\alpha^3} + \frac{y^{\alpha-\beta}\beta^3(\ln y)^3 \alpha^{(-3\beta-3)}}{1-y^{\alpha-\beta}} + \frac{3y^{\alpha-\beta}\beta^2(\beta+1)(\ln y)^2 \alpha^{(-2\beta-3)}}{1-y^{\alpha-\beta}}$$

$$+ \frac{3\beta^3 y^{2\alpha-\beta}(\ln y)^3 \alpha^{(-3\beta-3)}}{(1-y^{\alpha-\beta})^2} + \frac{y^{\alpha-\beta}\beta(\beta+1)(\beta+2)(\ln y) \alpha^{(-\beta-3)}}{1-y^{\alpha-\beta}}$$

$$+ \frac{3\beta^2(\beta+1) y^{2\alpha-\beta}(\ln y)^2 \alpha^{(-2\beta-3)}}{(1-y^{\alpha-\beta})^2} + \frac{2y^{3\alpha-\beta}\beta^3(\ln y)^3 \alpha^{(-3\beta-3)}}{(1-y^{\alpha-\beta})^3}$$

$$-2\beta(\beta+1)(\beta+2)\alpha^{(-\beta-3)}(\ln y) \quad \ldots \ldots \ldots \ldots \ldots \ldots \ldots \ldots \ldots \ldots \ldots (21)$$

$$\frac{\partial^3 l}{\partial \beta^3} = \frac{y^{\alpha-\beta}(\ln \alpha)^3(\ln y)\alpha^{(-\beta)}}{1-y^{\alpha-\beta}} + \frac{3y^{\alpha-\beta}(\ln \alpha)^3(\ln y)^2 \alpha^{(-2\beta)}}{1-y^{\alpha-\beta}}$$

$$+ \frac{y^{\alpha-\beta}(\ln \alpha)^3(\ln y)^3 \alpha^{(-3\beta)}}{1-y^{\alpha-\beta}} + \frac{3y^{2\alpha-\beta}(\ln \alpha)^3(\ln y)^2 \alpha^{(-2\beta)}}{(1-y^{\alpha-\beta})^2}$$

$$+ \frac{3y^{2\alpha-\beta}(\ln \alpha)^3(\ln y)^3 \alpha^{(-3\beta)}}{(1-y^{\alpha-\beta})^2} + \frac{2y^{3\alpha-\beta}(\ln \alpha)^3(\ln y)^3 \alpha^{(-3\beta)}}{(1-y^{\alpha-\beta})^3}$$

$$-2\alpha^{(-\beta)}(\ln y)(\ln \alpha)^3 \quad \ldots \ldots \ldots \ldots \ldots \ldots \ldots \ldots \ldots \ldots \ldots (22)$$

$$\frac{\partial^3 l}{\partial \beta \partial \alpha \partial \alpha} = \frac{1}{\alpha^2} - 2\beta(\beta+1)\alpha^{(-\beta-2)}(\ln y)(\ln \alpha) + 2(2\beta+1)\alpha^{(-\beta-2)}(\ln y)$$

$$+ \frac{y^{\alpha-\beta}\beta(2\beta+1)(\ln y)^2(\ln \alpha)\alpha^{(-2\beta-2)}}{1-y^{\alpha-\beta}} - \frac{y^{\alpha-\beta}\beta(\ln y)^2\alpha^{(-2\beta-2)}}{1-y^{\alpha-\beta}}$$

$$+ \frac{y^{\alpha-\beta}\beta^2(\ln y)^3(\ln \alpha)\alpha^{(-3\beta-2)}}{1-y^{\alpha-\beta}} + \frac{3y^{2\alpha-\beta}\beta^2(\ln y)^3(\ln \alpha)\alpha^{(-3\beta-2)}}{(1-y^{\alpha-\beta})^2}$$

$$- \frac{y^{\alpha-\beta}\beta(\ln y)^2\alpha^{(-2\beta-2)}}{1-y^{\alpha-\beta}} - \frac{y^{\alpha-\beta}(\beta+1)(\ln y)\alpha^{(-\beta-2)}}{1-y^{\alpha-\beta}}$$

$$- \frac{3y^{2\alpha-\beta}\beta(\ln y)^2 \alpha^{(-2\beta-2)}}{(1-y^{\alpha-\beta})^2} + \frac{y^{\alpha-\beta}\beta(1+\beta)(\ln \alpha)(\ln y)\alpha^{(-\beta-2)}}{1-y^{\alpha-\beta}}$$

$$+\frac{y^{\alpha^{-\beta}}\beta^2\,(\ln\alpha)\,(\ln y)^2 \alpha^{(-2\beta-2)}}{1-y^{\alpha^{-\beta}}} - \frac{y^{\alpha^{-\beta}}\beta\,(\ln y)\alpha^{(-\beta-2)}}{1-y^{\alpha^{-\beta}}}$$

$$+\frac{y^{2\alpha^{-\beta}}\beta^2\,(\ln y)^2\,(\ln\alpha)\,\alpha^{(-2\beta-2)}}{\left(1-y^{\alpha^{-\beta}}\right)^2} + \frac{\beta\,y^{2\alpha^{-\beta}}(2\beta+1)(\ln y)^2\,(\ln\alpha)\,\alpha^{(-2\beta-2)}}{\left(1-y^{\alpha^{-\beta}}\right)^2}$$

$$+\frac{2y^{3\alpha^{-\beta}}\beta^2\,(\ln\alpha)\,(\ln y)^3\,\alpha^{(-3\beta-2)}}{\left(1-y^{\alpha^{-\beta}}\right)^3} \quad\ldots\ldots\ldots\ldots\ldots\ldots\ldots\ldots\ldots\ldots (23)$$

$$\frac{\partial^3 l}{\partial\alpha\partial\beta\partial\beta} = 4\,\alpha^{(-\beta-1)}(\ln y)(\ln\alpha) - 2\beta\,\alpha^{(-\beta-1)}(\ln y)(\ln\alpha)^2$$

$$+\frac{\beta y^{\alpha^{-\beta}}\,(\ln\alpha)^2\,(\ln y)^3\,\alpha^{(-3\beta-1)}}{1-y^{\alpha^{-\beta}}} - \frac{2y^{\alpha^{-\beta}}(\ln y)^2\,(\ln\alpha)\,\alpha^{(-2\beta-1)}}{1-y^{\alpha^{-\beta}}}$$

$$+\frac{3y^{\alpha^{-\beta}}\beta(\ln y)^2\,(\ln\alpha)^2\,\alpha^{(-2\beta-1)}}{1-y^{\alpha^{-\beta}}} + \frac{3y^{2\alpha^{-\beta}}\beta\,(\ln y)^3\,(\ln\alpha)^2\,\alpha^{(-3\beta-1)}}{\left(1-y^{\alpha^{-\beta}}\right)^2}$$

$$-\frac{2y^{\alpha^{-\beta}}\,(\ln\alpha)\,(\ln y)\,\alpha^{(-\beta-1)}}{1-y^{\alpha^{-\beta}}} - \frac{2y^{2\alpha^{-\beta}}\beta\,(\ln y)^2\,(\ln\alpha)\,\alpha^{(-2\beta-1)}}{\left(1-y^{\alpha^{-\beta}}\right)^2}$$

$$+\frac{y^{\alpha^{-\beta}}\beta\,(\ln y)\,(\ln\alpha)^2 \alpha^{(-\beta-1)}}{1-y^{\alpha^{-\beta}}} + \frac{3y^{2\alpha^{-\beta}}\beta\,(\ln y)^2\,(\ln\alpha)^2\,\alpha^{(-2\beta-1)}}{\left(1-y^{\alpha^{-\beta}}\right)^2}$$

$$+\frac{2\,\beta y^{3\alpha^{-\beta}}\,(\ln\alpha)^2\,(\ln y)^3\,\alpha^{(-3\beta-1)}}{\left(1-y^{\alpha^{-\beta}}\right)^3} \quad\ldots\ldots\ldots\ldots\ldots\ldots\ldots\ldots\ldots\ldots (24)$$

Taking expectation of the above derivative is carried out using Monte Carlo integration; see equations (25-31). The integrals in these equations can be calculated using the appropriate method of numerical integration and it is a fixed number. The author underwent trapezoid method for integration. Substituting the estimated parameters alpha and beta for each data set, the following integrals can be calculated and are fixed value for each dataset.

$$E\left(\frac{y^{\alpha^{-\beta}}\,(\ln y)^3}{1-y^{\alpha^{-\beta}}}\right) = \int_0^1 \frac{y^{\alpha^{-\beta}}\,(\ln y)^3}{1-y^{\alpha^{-\beta}}} f(y)dy = f_1 \quad\ldots\ldots\ldots\ldots\ldots\ldots (25)$$

$$E\left(\frac{y^{\alpha^{-\beta}}\,(\ln y)^2}{1-y^{\alpha^{-\beta}}}\right) = \int_0^1 \frac{y^{\alpha^{-\beta}}\,(\ln y)^2}{1-y^{\alpha^{-\beta}}} f(y)dy = f_2 \quad\ldots\ldots\ldots\ldots\ldots\ldots (26)$$

$$E\left(\frac{y^{\alpha-\beta} \ln y}{1-y^{\alpha-\beta}}\right) = \int_0^1 \frac{y^{\alpha-\beta} (\ln y)}{1-y^{\alpha-\beta}} f(y)dy = f_3 \quad \ldots\ldots\ldots\ldots\ldots\ldots\ldots\ldots\ldots\ldots (27)$$

$$E\left(\frac{y^{2\alpha-\beta} (\ln y)^3}{[1-y^{\alpha-\beta}]^2}\right) = \int_0^1 \frac{y^{2\alpha-\beta} (\ln y)^3}{[1-y^{\alpha-\beta}]^2} f(y)dy = f_4 \quad \ldots\ldots\ldots\ldots\ldots\ldots\ldots\ldots (28)$$

$$E\left(\frac{y^{2\alpha-\beta} (\ln y)^2}{[1-y^{\alpha-\beta}]^2}\right) = \int_0^1 \frac{y^{2\alpha-\beta} (\ln y)^2}{[1-y^{\alpha-\beta}]^2} f(y)dy = f_5 \quad \ldots\ldots\ldots\ldots\ldots\ldots\ldots\ldots (29)$$

$$E\left(\frac{y^{3\alpha-\beta} (\ln y)^3}{[1-y^{\alpha-\beta}]^3}\right) = \int_0^1 \frac{y^{3\alpha-\beta} (\ln y)^3}{[1-y^{\alpha-\beta}]^3} f(y)dy = f_6 \quad \ldots\ldots\ldots\ldots\ldots\ldots\ldots\ldots (30)$$

$$E(\ln y) = \int_0^1 (\ln y) f(y)dy = f_7 \quad \ldots\ldots\ldots\ldots\ldots\ldots\ldots\ldots\ldots\ldots\ldots\ldots\ldots (31)$$

Define the following quantities:

$$k_{11} = E\left(\frac{\partial^2 l}{\partial \alpha^2}\right), k_{22} = E\left(\frac{\partial^2 l}{\partial \beta^2}\right), \quad k_{12} = E\left(\frac{\partial^2 l}{\partial \alpha \partial \beta}\right) = k_{21} = E\left(\frac{\partial^2 l}{\partial \beta \partial \alpha}\right)$$

$$k_{111} = E\left(\frac{\partial^3 l}{\partial \alpha^3}\right), \quad k_{222} = E\left(\frac{\partial^3 l}{\partial \beta^3}\right), k_{121} = E\left(\frac{\partial^3 l}{\partial \alpha \partial \beta \partial \alpha}\right) = k_{211} = E\left(\frac{\partial^3 l}{\partial \beta \partial \alpha \partial \alpha}\right)$$

$$k_{122} = E\left(\frac{\partial^3 l}{\partial \alpha \partial \beta \partial \beta}\right) = k_{212} = E\left(\frac{\partial^3 l}{\partial \beta \partial \alpha \partial \beta}\right)$$

$$k_{112} = E\left(\frac{\partial^3 l}{\partial \alpha^2 \partial \beta}\right), \quad k_{221} = E\left(\frac{\partial^3 l}{\partial \beta^2 \partial \alpha}\right)$$

$$k_{11}^{(1)} = \frac{\partial k_{11}}{\partial \alpha}, \quad k_{12}^{(1)} = \frac{\partial k_{12}}{\partial \alpha} = k_{21}^{(1)} = \frac{\partial k_{21}}{\partial \alpha}, \quad k_{22}^{(1)} = \frac{\partial k_{22}}{\partial \alpha}$$

$$k_{11}^{(2)} = \frac{\partial k_{11}}{\partial \beta}, \quad k_{12}^{(2)} = \frac{\partial k_{12}}{\partial \beta} = k_{21}^{(2)} = \frac{\partial k_{21}}{\partial \beta}, \quad k_{22}^{(2)} = \frac{\partial k_{22}}{\partial \beta}$$

$$a_{11}^{(1)} = k_{11}^{(1)} - 0.5\, k_{111}, \quad a_{12}^{(1)} = k_{12}^{(1)} - 0.5\, k_{121} \quad , \quad a_{22}^{(1)} = k_{22}^{(1)} - 0.5\, k_{221}$$

$$a_{11}^{(2)} = k_{11}^{(2)} - 0.5\, k_{112}, \quad a_{12}^{(2)} = k_{12}^{(2)} - 0.5\, k_{122} \quad , \quad a_{22}^{(2)} = k_{22}^{(2)} - 0.5\, k_{222}$$

The information matrix is $K = \{-k_{ij}\} = -n \times \begin{bmatrix} k_{11} & k_{12} \\ k_{21} & k_{22} \end{bmatrix}$, where n is the number of observations.

Defining $A_{ij}^{(q)} = \{a_{ij}^{(q)}\}; q = 1,2$ and $A = [A^{(1)} \mid A^{(2)}]$

$$A = n \times \begin{bmatrix} a_{11}^{(1)} & a_{12}^{(1)} & a_{11}^{(2)} & a_{12}^{(2)} \\ a_{21}^{(1)} & a_{22}^{(1)} & a_{21}^{(2)} & a_{22}^{(2)} \end{bmatrix}$$

Upon using Cordeiro and Klein (1994) modification of the Cox and Snell (1968) result; the $Bias\begin{pmatrix}\hat{\alpha}\\\hat{\beta}\end{pmatrix} = K^{-1} A \, vec\,(K^{-1})$

The bias-adjusted estimators can be obtained as in equation (32)

$$\begin{pmatrix}\alpha^*\\\beta^*\end{pmatrix} = \hat{K}^{-1}\hat{A} \, vec\,(\hat{K}^{-1}) \ldots\ldots\ldots\ldots\ldots\ldots\ldots\ldots\ldots\ldots\ldots\ldots\ldots\ldots\ldots\ldots\ldots\ldots\ldots\ldots\ldots(32)$$

Where $\hat{K} = K|_{\alpha=\hat{\alpha} \,\&\, \beta=\hat{\beta}}$ and $\hat{A} = A|_{\alpha=\hat{\alpha} \,\&\, \beta=\hat{\beta}}$

# Results

## Section 3

### 3.1. Variance-corrected MLE procedure and MBUW

Maximum likelihood estimators for the parameters of MBUW distribution may have large variances or infinite variances. Because the parameters are correlated they also have large covariance. The surface of the log-likelihood function may exhibit a flat shape which is reflected by multiple pair of estimators that fit the distribution. Also the estimators may be unstable. Inspecting the surface of the negative log-likelihood function can depict the pairs of parameters which attain the minimal negative likelihood (nLL) values. The approach used by the author in this paper is to obtain these pairs and find the relationship between them and define one parameter in term of the other. Then re-parameterize the negative likelihood function and scale one of the parameter in log scale if needed. Then estimate this parameter and back substitute in the relationship to obtain the other parameter. The variance of the parameter obtained from the MLE is obtained by inversing the fisher information while the parameter defined in the relationship is obtained using the delta method.

### 3.2. Real data analysis using variance-corrected MLE

These data were mentioned by the author in previous work (Iman M. Attia, 2024)

Steps of the technique used by the author:

1. Inspect the surface of the nLL.
2. Extract the pairs of alpha and beta parameter that minimize the nLL. The range of the alpha and beta were divided into equally spaced 500 points. The range differs according to the datasets. These ranges were chosen according to the results obtained from MLE using Nelder Mead algorithm results.
3. Define beta in terms of alpha by fitting the best curve for this relationship.
4. Re-parameterize the nLL using this relationship.
5. Estimate alpha parameter.
6. Back substitute in the relationship to obtain the beta parameter.
7. Use inverse fisher obtained from the MLE to obtain variance of the alpha and use the delta method to obtain the variance of the beta.
8. Use goodness of fit test like the KS-test, AD-test and CVM-test to evaluate the fitting of the distribution to data.

### 3.2.1. First dataset: Dwelling without basic facilities.

1. Inspecting the surface of nLL. Figure 1 illustrates this surface of which has a flat part.
2. Extracting the pairs of the parameters that minimize the nLL. These pairs were 130 pairs. Figure 2 illustrates the relation between these parameters.
3. Define the relationship between alpha and beta by fitting the best curve depicting this relationship. see equations (33-35)

   For exponential decay model: $\beta = 5.4726\, e^{-0.7058} + 0.845 \ldots\ldots\ldots\ldots\ldots. (33)$

   For polynomial model: $\beta = 0.0618\, \alpha^2 - 0.7674\, \alpha + 3.279 \ldots\ldots\ldots\ldots\ldots. (34)$

   For reciprocal model: $\beta = \dfrac{3.6607}{\alpha} + 0.2907 \ldots\ldots\ldots\ldots\ldots\ldots\ldots\ldots\ldots (35)$

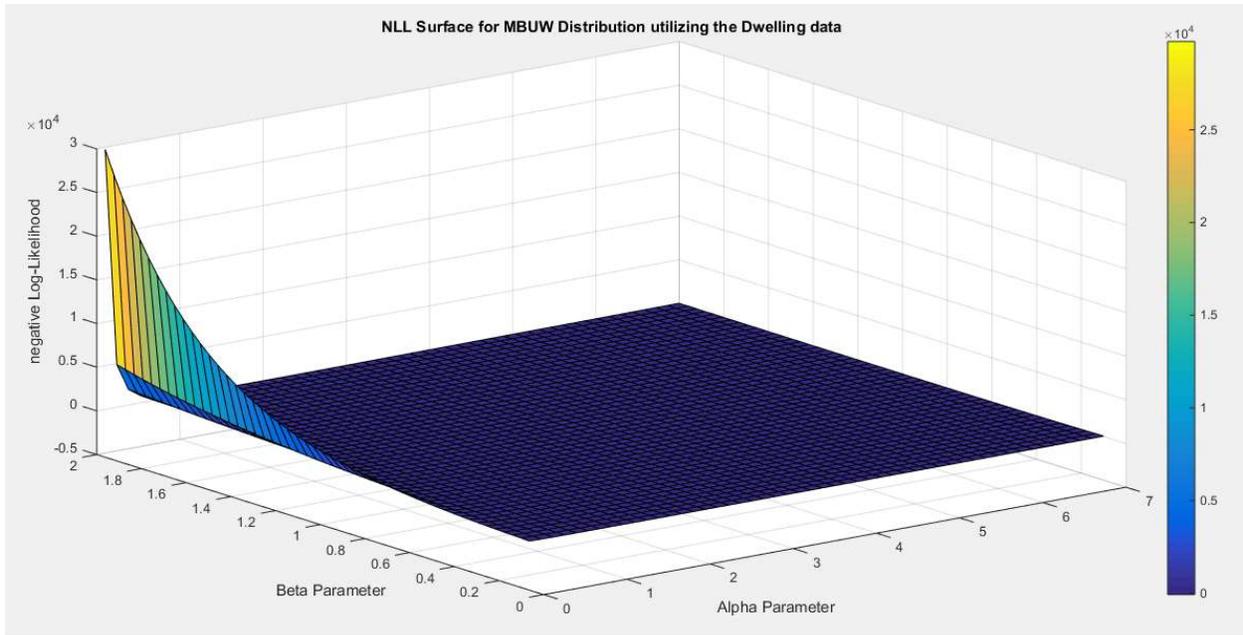

Fig. 1 negative likelihood surface using the dwelling data.

4- Re-parameterize the nLL by substituting each beta in LL as in equation (36)

$$l(\alpha) = n\ln(6) - n\beta\ln(\alpha) + \sum_{i=1}^{n} \ln\left[1 - y_i^{\frac{1}{\alpha^\beta}}\right] + \left(\frac{2}{\alpha^\beta} - 1\right)\sum_{i=1}^{n} \ln(y_i) \ldots \ldots \ldots (36)$$

5- Use the Nelder Mead algorithm to obtain MLE estimator for the alpha and use the inverse fisher to obtain the variance of the alpha.
6- Back substitute in each model to obtain the beta.
7- Use the delta method to obtain the variance of the beta as follows: delta method is defined as $var\left(g(\hat{\theta})\right) = \left(g'(\theta)\right)^2 var(\hat{\theta})$ as in equations(37-40)

For the three models it is defined as: $var(\hat{\beta}) = \left[\frac{d}{d\alpha}(\beta)\right]^2 var(\hat{\alpha}) \ldots \ldots \ldots \ldots (37)$

For the exponen. Decay : $\frac{d}{d\alpha}\beta = 5.4726(-0.7058)\, e^{-0.7058} \ldots \ldots \ldots \ldots \ldots \ldots (38)$

For the quadratic polynomial: $\frac{d}{d\alpha}\beta = 2*0.0618\,\alpha - 0.7674 \ldots \ldots \ldots \ldots \ldots (39)$

For the reciprocal model $\frac{d}{d\alpha}\beta = \frac{-3.6607}{\alpha^2} \ldots \ldots \ldots \ldots \ldots \ldots \ldots \ldots \ldots \ldots (40)$

8- Use goodness of fit techniques such as KS-test to test the null hypothesis testing the fitness of the data to the theoretical distribution with the estimated parameters.

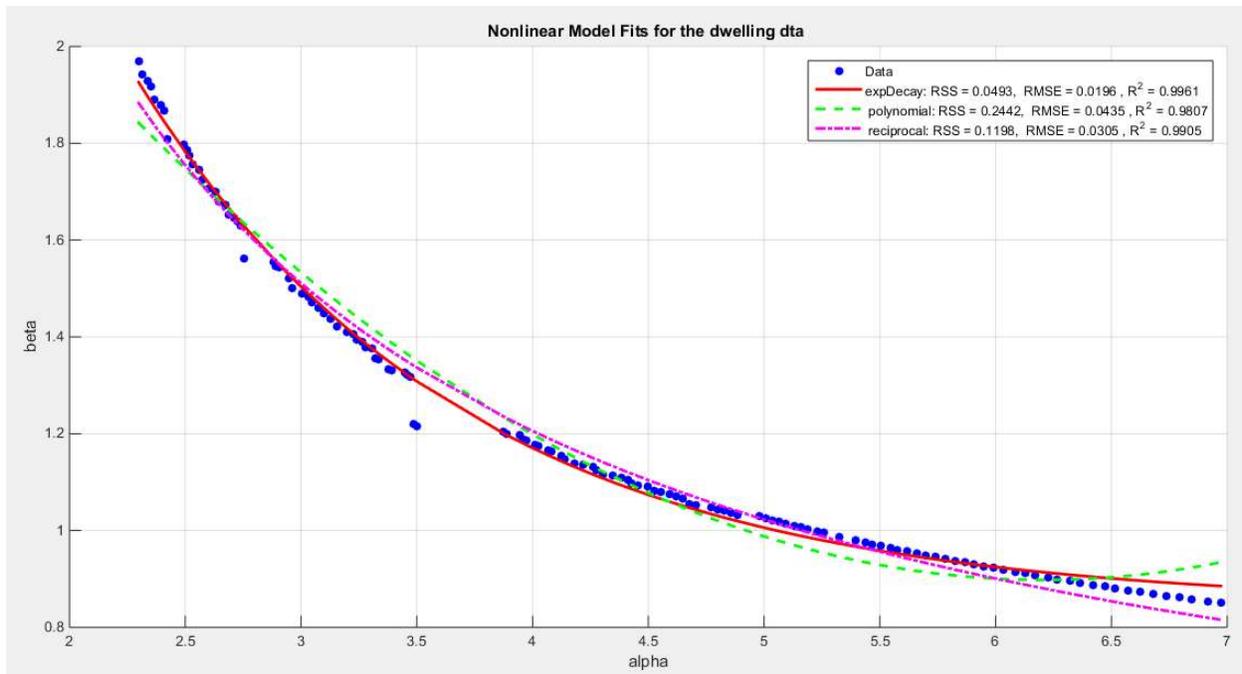

Fig. 2 shows a decreasing convex relationship between alpha and beta. The Nonlinear models for the dwelling dataset, with residual sum of squares (RSS), R^2 and root of mean square error(RMSE) are illustrated in the figure for each curve, high lightening that the exponential decay is the best model followed by the reciprocal model then the polynomial (quadratic) model.

The results of the above procedure are shown in Table 1

The exponential decay has the following metrics: residual sum of square (RSS) value 0.0493, R sqaure value 0.9961, and root mean square error (RMSE) value 0.0195. The reciprocal model has the following values for RSS, R squared and, RMSE: 0.1198, 0.9905 and, 0.0305 respectively. The quadratic model has the following values for RSS, R squared and, RMSE: 0.2442, 0.9807 and 0.0435 respectively.

The analysis revealed marked reduction in the variance for both alpha and beta. This has tremendous effect when constructing the confidence interval (CI) as it becomes narrower. The values of the parameters were not unique; they show small changes when changing the initial guess. The values with minimum variance were chosen to be reported as long as they failed to reject the null hypothesis.  The other metrics like nLL, information Akiake information criteria (AIC), corrected AIC (CAIC), Bayesian information criteria (BIC), and Hannon Quin information criteria (HQIC) are the same as the one that are obtained from conducting Nelder mead algorithm to estimate MLE for both parameters. And this is reflected on the graph of the theoretical CDF, graph of PDF and the QQ plot. They are all the same as  there is a same pattern if the parameters give the same nLL. The values of alpha and beta are nearly equal across the models and the variances show minor differences among the models but the quadratic model has the minimal variance.

Table 1: the results of the 3 nonlinear model using the dwelling data

| metric | Exponential decay | Quadratic-polynomial | Reciprocal |
|---|---|---|---|
| Alpha | 2.7988 (2.7286,2.869) | 2.7268 (2.6602,2.7935) | 2.8707 (2.7969,2.9444) |
| Beta | 1.6045 (1.5669,1.6421) | 1.6461 (1.6175,1.6748) | 1.5659 (1.5331,1.5987) |
| Var (SE) alpha | 0.0449 (0.0358) | 0.0405 (0.034) | 0.0496 (0.0376) |
| Var (SE) beta | 0.0129 (0.0192) | 0.0075 (0.0146) | 0.0098 (0.0167) |
| nLL | -74.2925 | -74.2925 | -74.2925 |
| KS | 0.1794 | 0.1794 | 0.1794 |
| AD | 2.3889 | 2.3889 | 2.3889 |
| CVM | 0.3966 | 0.3966 | 0.3966 |
| AIC | -144.585 | -144.585 | -144.585 |
| CAIC | -144.21 | -144.21 | -144.21 |
| BIC | -141.4743 | -141.4743 | -141.4743 |
| HQIC | -3.5422 | -3.5422 | -3.5422 |
| Ho | Fail to reject | Fail to reject | Fail to reject |
| P-value (KS-test) | 0.186 | 0.186 | 0.186 |

## 3.2.2. Second dataset: Quality of the support network

1- Inspecting the surface of nLL. Figure 3 illustrates this surface of which has a flat part.
2- Extracting the pairs of the parameters that minimize the nLL. These pairs were 16 pairs. Figure 4 illustrates the relation between these parameters.
3- Define the relationship between alpha and beta by fitting the best curve depicting this relationship. See equation (41)
   For linear model: $\beta = 4.2007\, \alpha + 0.442 \dots \dots \dots \dots \dots \dots \dots \dots \dots \dots \dots \dots \dots \dots \dots \dots$ (41)

Follow the steps from 4 to 8 and Table (2-3) shows the results.

The linear model has the following values for RSS, R squared and, RMSE: 0.0029, 0.9977 and, 0.014 respectively.

Table 2 results from second dataset

| Alpha & CI | Beta & CI | Var(alpha) & SE | Var(beta) & SE | $H_0$&p(KS) |
|---|---|---|---|---|
| 0.0558 (0.0492,0.0624) | 0.6763 (0.6486,0.704) | 0.00023 (0.0034) | 0.004 (0.0141) | Fail to reject (p=0.8235) |

Table 3 results from second dataset (continuation)

| AIC | CAIC | BIC | HQIC | KS | AD | CVM | nLL |
|---|---|---|---|---|---|---|---|
| -55.734 | -55.0281 | -53.7425 | -2.4048 | 0.1061 | 0.452 | 0.0803 | -29.867 |

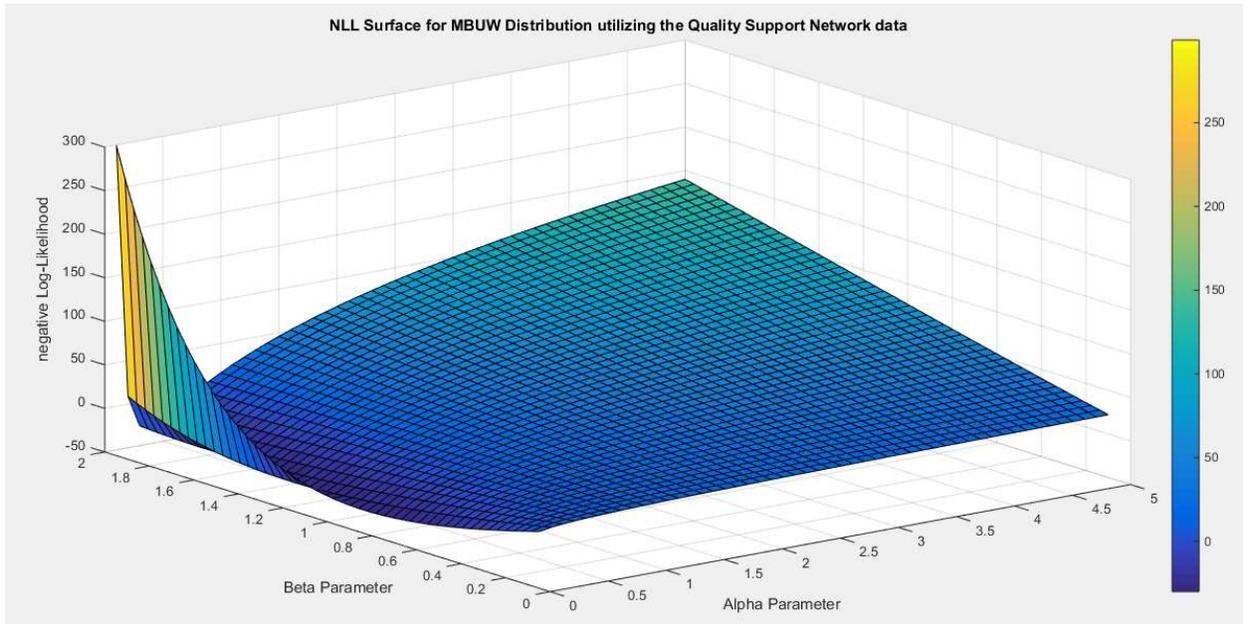

Fig. 3 negative likelihood surface using the Quality of support network dataset.

The variance of both parameters show marked reduction. The confidence (CI) interval becomes narrower. The distribution fits the data. Other metrics are more or less the same as the results obtained from MLE using the Nelder Mead algorithm. The parameters do not change with the initial guess.

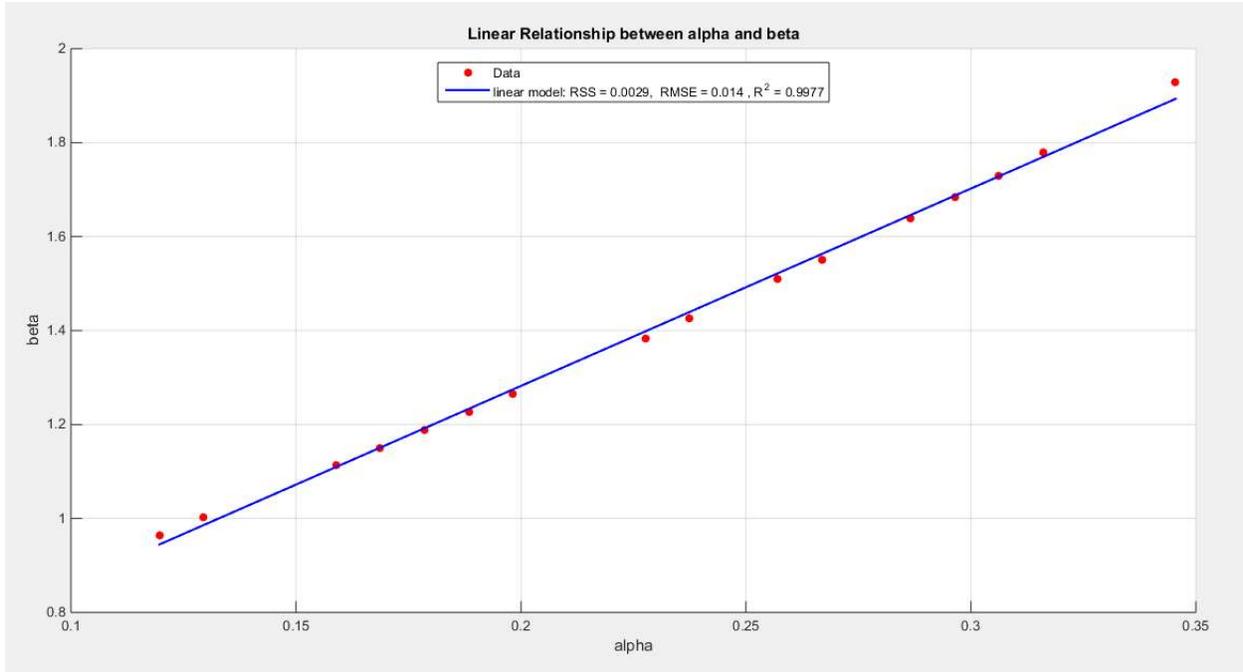

Fig. 4 shows the increasing linear relationship between the alpha and the beta parameter for the quality of support networks, with RSS=0.0029, R^2=0.9977 and RMSE=0.014 shown in the figure.

### 3.2.3. Third dataset: voter turnout dataset

1- Inspecting the surface of nLL. Figure 5 illustrates this surface of which has a flat part.
2- Extracting the pairs of the parameters that minimize the nLL. These pairs were 51 pairs. Figure 6 illustrates the relation between these parameters.
3- Define the relationship between alpha and beta by fitting the best curve depicting this relationship. See equations (42-43)

For quadratic model: $\beta = 4.3363\, \alpha^2 - 0.86455\, \alpha + 0.47298$ … … … … … … … … (42)

For the power law model: $\beta = 3.8879\, \alpha^{2.4616} + 0.4082$ … … … … … … … … … .. (43)

Follow the steps from 4 to 8, for the delta method see equation (44-45):

For the power law: $\frac{d}{d\alpha}\beta = 3.8879\, (2.4616)\, \alpha^{1.4616}$ … … … … … … … … … … … . . (44)

For the quadratic polynomial: $\frac{d}{d\alpha}\beta = 2 * 4.3363\, \alpha - 0.8645$ … … … … … … … . . (45)

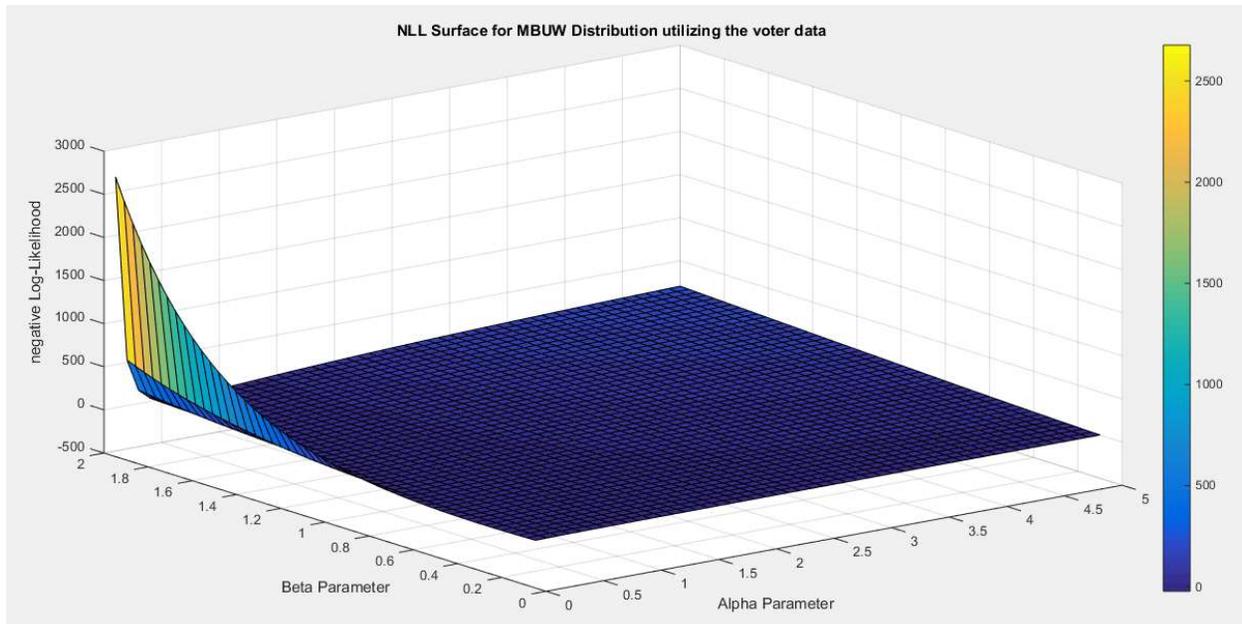

Fig.5 negative likelihood surface using the Voter dataset.

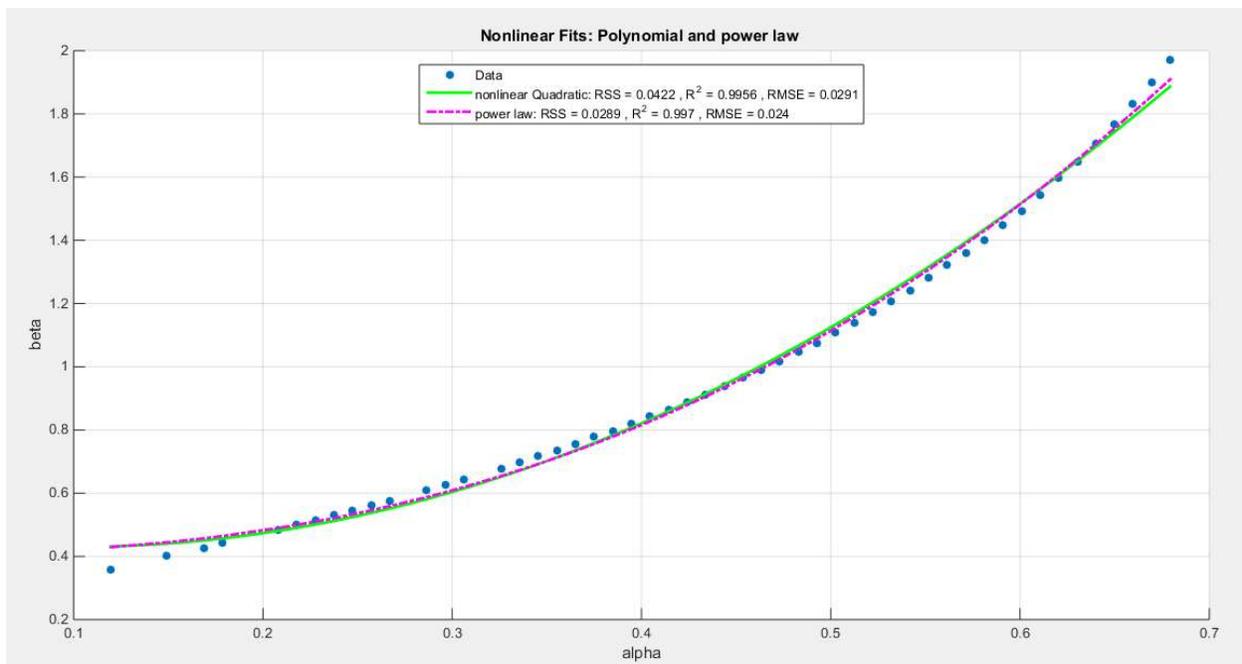

Fig. 6 Nonlinear models for the voter dataset (increasing convex relationship between alpha and the beta), with residual sum of squares (RSS), R^2 and root of mean square error(RMSE) illustrated in the figure for each curve, high lightening that the power law is the best model followed by the polynomial (quadratic).

The power law model has the following values for RSS, R squared and RMSE: 0.0289, 0.997, and 0.024 respectively. The quadratic model has the following values for RSS, R squared and RMSE: 0.0422, 0.9956, and 0.0291 respectively.

Table 4 illustrates the results of using these models to re-parameterize the nLL and running MLE using the Nelder Mead algorithm by Matlab.

From Table 4, the analysis revealed that the metrics between the two models are minor but it is in favor of the polynomial model over the power law model. This is due to the fact that it has more negative values of the nLL, AIC, CAIC, BIC, HQIC than the power law has. It also has less value for the KS-test, AD and CVM than the power law has. All these metrics favor the quadratic model over the power law model, although the power law model better describes the relationship between beta and alpha than the quadratic model as indicated by the metrics; SRR, R squared and RMSE.

Table 4: the results of the 2 nonlinear models using the voter data

| metric | Power law model | Quadratic-polynomial model |
|---|---|---|
| Alpha(CI) | 0.483 (0.4661,0.4998) | 0.4297 (0.4118,0.4471) |
| Beta (CI) | 1.0563 (1.0007,1.112) | 0.9014 (0.8509,0.9519) |
| Var (SE)alpha | 0.0028(0.0086) | 0.0031(0.009) |
| Var(SE)beta | 0.0306(0.0284) | 0.0252(0.0258) |
| nLL | -22.0671 | -22.0688 |
| KS | 0.1395 | 0.1364 |
| AD | 1.3657 | 1.3262 |
| CVM | 0.2295 | 0.2193 |
| AIC | -40.1342 | -40.1377 |
| CAIC | -39.7914 | -39.7948 |
| BIC | -36.859 | -36.8625 |
| HQIC | -1.0229 | -1.0231 |
| H₀ = 0 | Fail to reject | Fail to reject |
| P-value(KS-test) | 0.412 | 0.4401 |

### 3.2.4. Fourth dataset: Flood data.

1- Inspecting the surface of nLL. Figure 7 illustrates this surface of which has a flat part.
2- Extracting the pairs of the parameters that minimize the nLL. These pairs were 139 pairs. Figure 8 illustrates the relation between these parameters.
3- Define the relationship between alpha and beta by fitting the best curve depicting this relationship. See equations (46-48)

For exponentional decay model: $\beta = 19361\, e^{-9.1127} + 0.1435$ ............... (46)

For polynomial model: $\beta = 0.889\, \alpha^2 - 3.5278\, \alpha + 3.5557$ ................. (47)

For reciprocal model: $\beta = \frac{1.3445}{\alpha} - 0.5694$ ........................ (48)

Follow the steps from 4 to 8

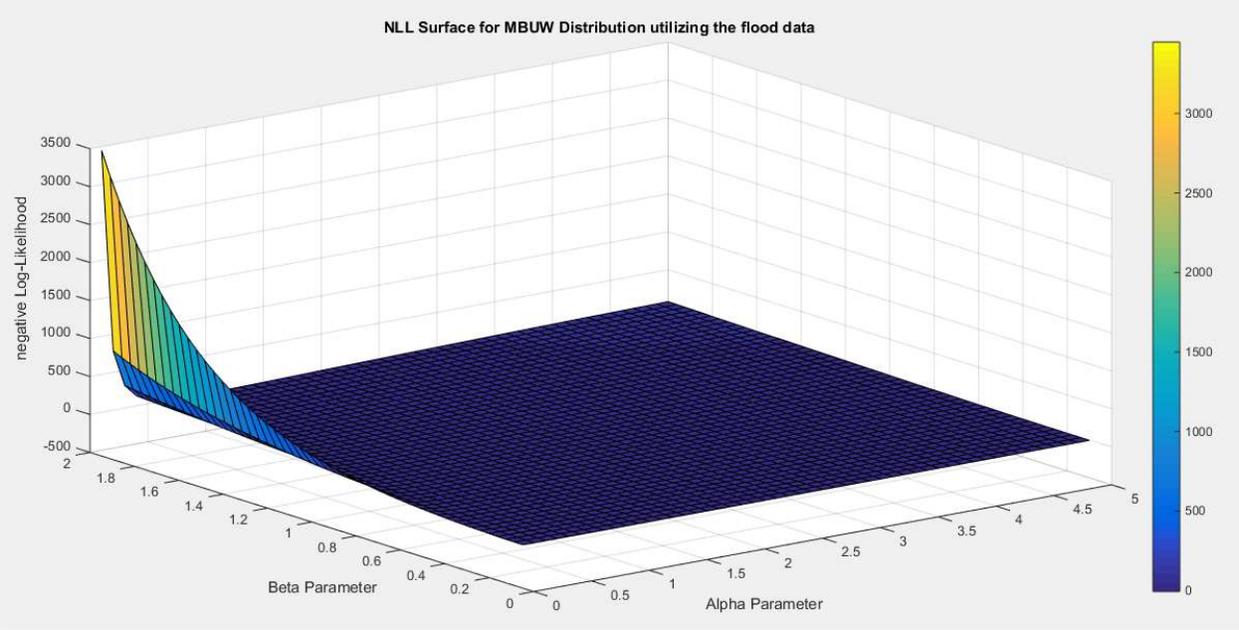

Fig. 7 negative likelihood surface using the flood dataset.

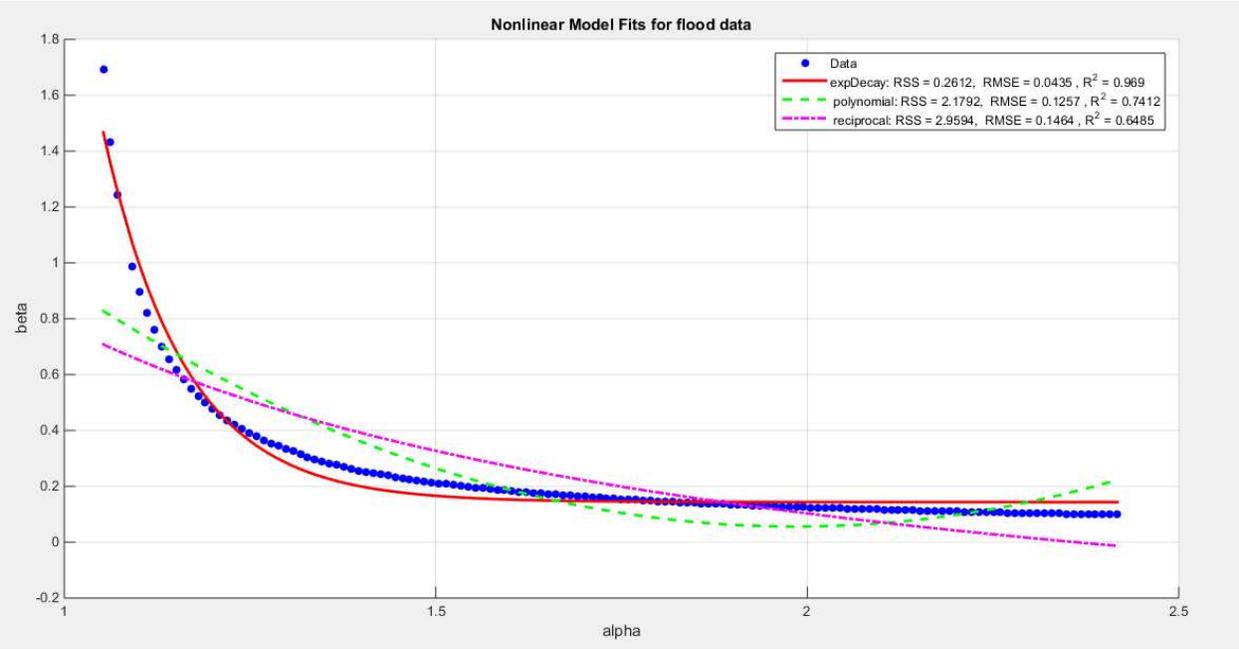

Fig. 8 shows a decreasing convex relationship between alpha and beta. The Nonlinear models for the flood dataset, with residual sum of squares (RSS), R^2 and root of mean square error(RMSE) are illustrated in the figure for each curve, high lightening that the exponenetial is the best model followed by the polynomial( quadratic) then the reciprocal model.

The exponential decay model has the following values for RSS, R squared and, RMSE: 0.2612, 0.969, and 0.0435 respectively. The quadratic model has the following values for RSS, R squared and, RMSE: 2.1792, 0.7412 and 0.1257 respectively. The reciprocal model has the following values for RSS, R squared and, RMSE: 2.9594, 0.6485 and, 0.1464 respectively.

Table 5 illustrates the results obtained from re-parameterizing the nLL function using these 3 different models.

In the analysis, initial guess may cause the estimators to change their values, but the variance and the goodness of fit may judge or add to the control if this initial guess, causing the new value, should be taking into consideration or discarded. In this situation the low variance estimator associated with goodness of fit test supporting the fitting distribution are the corner stones to choose the estimators values and hence consider the stability of the estimators.

Table 5: the results of the 3 nonlinear models using the flood data

| metric | Exponential decay | Quadratic-polynomial | Reciprocal |
|---|---|---|---|
| Alpha | 1.0712 | 1.1312 | 1.1575 |
|  | (1.0114 , 1.1311) | (1.0179 , 1.2444) | (1.02 , 1.295) |
| Beta | 1.2591 | 0.7027 | 0.5921 |
|  | (0.6507,1.8675) | (0.5309,0.8744) | (0.4542,0.7301) |
| Var(SE) alpha | 0.0186 | 0.0668 | 0.0984 |
|  | (0.0305) | (0.0578) | (0.0702) |
| Var(SE) beta | 1.9269 | 0.1535 | 0.0991 |
|  | (0.3104) | (0.0876) | (0.0704) |
| nLL | -6.4617 | -6.4617 | -6.4617 |
| KS | 0.3202 | 0.3202 | 0.3202 |
| AD | 2.7563 | 2.7563 | 2.7563 |
| CVM | 0.531 | 0.531 | 0.531 |
| AIC | -8.9233 | -8.9233 | -8.9233 |
| CAIC | -8.2174 | -8.2174 | -8.2174 |
| BIC | -6.9319 | -6.9319 | -6.9319 |
| HQIC | 0.657 | 0.657 | 0.657 |
| Ho | Fail to reject | Fail to reject | Fail to reject |
| P-value (KS-test) | 0.0253 | 0.0253 | 0.0253 |

The variance of alpha obtained by exponential decay model is the lowest than other variances obtained by quadratic and reciprocal model. The opposite is true concerning the variance of the beta. The variances of both alpha and beta for the reciprocal model are nearly equal as well as the standard errors. This is reflected by a narrower CI for both alpha and beta estimators obtained from the reciprocal model. This is not true concerning the interval for the beta estimator obtained from exponential decay model which is the widest among the other CI intervals obtained from other models. So the reciprocal model may outperform the exponential model

## 3.2.5. Fifth dataset: time between failures data set.

1. Inspecting the surface of nLL. Figure 9 illustrates this surface of which has a flat part.
2. Extracting the pairs of the parameters that minimize the nLL. These pairs were 246 pairs. Figure 10 illustrates the relation between these parameters.
3. Define the relationship between alpha and beta by fitting the best curve depicting this relationship. See equations (49-51)

   For exponential decay model: $\beta = 7.1594\, e^{-0.97948} + 0.6735$ ……………….. (49)

   For polynomial model: $\beta = 0.1011\, \alpha^2 - 1.0218\, \alpha + 3.2653$ ………………. (50)

   For reciprocal model: $\beta = \frac{3.1318}{\alpha} + 0.0648$ …………………………….. (51)

   Follow the steps from 4 to 8

The exponential decay model has the following values for RSS, R squared and, RMSE: 0.1239, 0.9957, and 0.0225 respectively. The reciprocal model has the following values for RSS, R squared and, RMSE: 0.5583, 0.9808 and, 0.0477 respectively. The quadratic model has the following values for RSS, R squared and, RMSE: 0.9653, 0.9668 and 0.0628 respectively. Table 6 illustrates the results obtained from re-parameterizing the nLL function using these 3 different models.

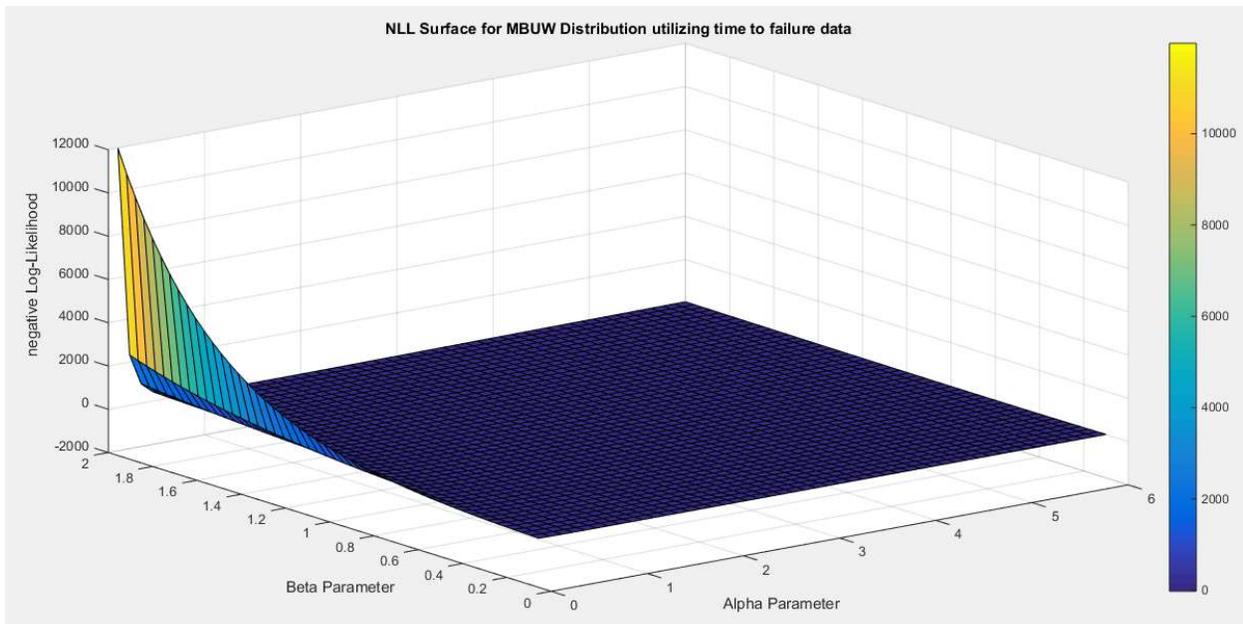

Fig. 9 negative likelihood surface using the time between failures dataset.

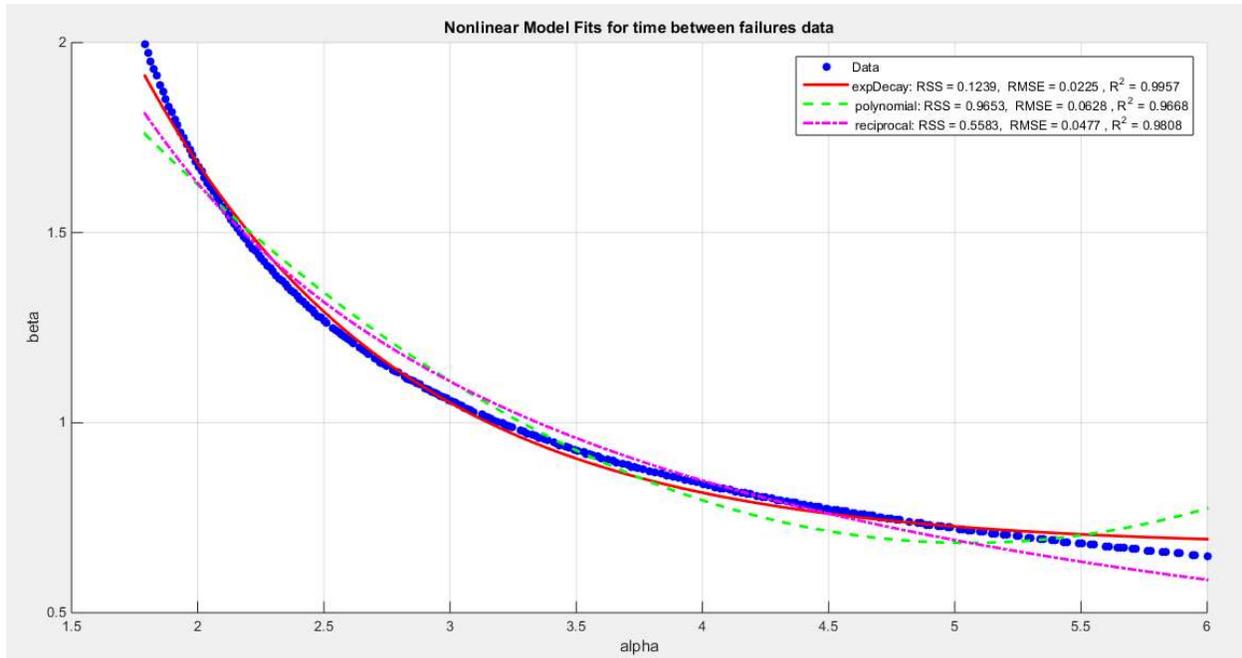

Fig. 10 shows a decreasing convex relationship between alpha and beta. The Nonlinear models for the time between failures dataset, with residual sum of squares (RSS), R^2 and root of mean square error(RMSE) are illustrated in the figure for each curve, high lightening that the exponenetial is the best model followed by the reciprocal model then the polynomial (quadratic).

Table 6: the results of the 3 nonlinear models using the time between failures data

| metric | Exponential decay | Quadratic-polynomial | Reciprocal |
|---|---|---|---|
| Alpha | 1.977 (1.906 , 2.048 ) | 2.105 (2.0224 , 2.1875) | 2.141 (2.0551 , 2.2269 ) |
| Beta | 1.7062 (1.6343,1.778) | 1.5624 (1.5132 , 1.6116 ) | 1.5276 (1.4689 , 1.5853 ) |
| Var (SE) alpha | 0.0302 (0.0362) | 0.0408 (0.0421) | 0.0442 (0.0438) |
| Var (SE) beta | 0.0309 (0.0367) | 0.0145 (0.0251) | 0.0206 (0.0299) |
| nLL | -19.931 | -19.931 | -19.931 |
| KS | 0.1584 | 0.1584 | 0.1584 |
| AD | 0.6703 | 0.6703 | 0.6703 |
| CVM | 0.1253 | 0.1253 | 0.1253 |
| AIC | -35.862 | -35.862 | -35.862 |
| CAIC | -35.262 | -35.262 | -35.262 |
| BIC | -33.591 | -33.591 | -33.591 |
| HQIC | -1.4134 | -1.4134 | -1.4134 |
| $H_o = 0$ | Fail to reject | Fail to reject | Fail to reject |
| P-value (KS-test) | 0.5575 | 0.5575 | 0.5575 |

The analysis of the above models is comparable the variances are approximately equal, the estimators are also nearly equal. Any model can be chosen. The CI is narrow. The estimators are stable regardless the initial guess used as the estimator, variance and the goodness of fit results does not change. This is peculiar for this dataset.

### 3.2.6. Sixth dataset: capacity factor data set.

1. Inspecting the surface of nLL. Figure 11 illustrates this surface of which has a flat part.
2. Extracting the pairs of the parameters that minimize the nLL. These pairs were 80 pairs. Figure 12 illustrates the relation between these parameters.
3. Define the relationship between alpha and beta by fitting the best curve depicting this relationship. See equation (52-54)
   For exponential decay model: $\beta = 9.424\, e^{-1.2276} + 0.6253$ ……………………. (52)

   For polynomial model: $\beta = 0.1596\, \alpha^2 - 1.3633\, \alpha + 3.537$ ………………….. (53)

   For reciprocal model: $\beta = \frac{3.0101}{\alpha} - 0.0811$ ………………………………… (54)

   Follow the steps from 4 to 8

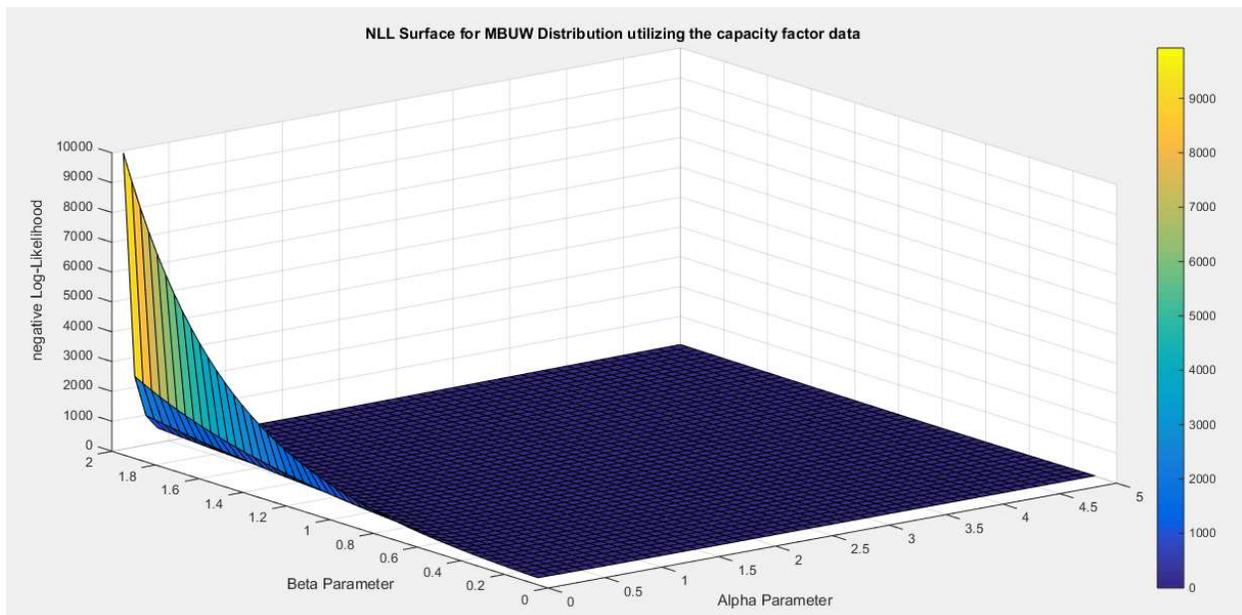

Fig. 11 negative likelihood surface using the factors affecting the unit capacity dataset.

The exponential decay model has the following values for RSS, R squared and, RMSE: 0.0442, 0.9954, and 0.0237 respectively. The reciprocal model has the following values for RSS, R squared and, RMSE: 0.2843, 0.9701 and 0.06 respectively. The quadratic model has the following values for RSS, R squared and, RMSE: 0.3523, 0.963 and 0.0668 respectively.

Table 7 illustrates the results obtained from re-parameterizing the nLL function using these 3 different models.

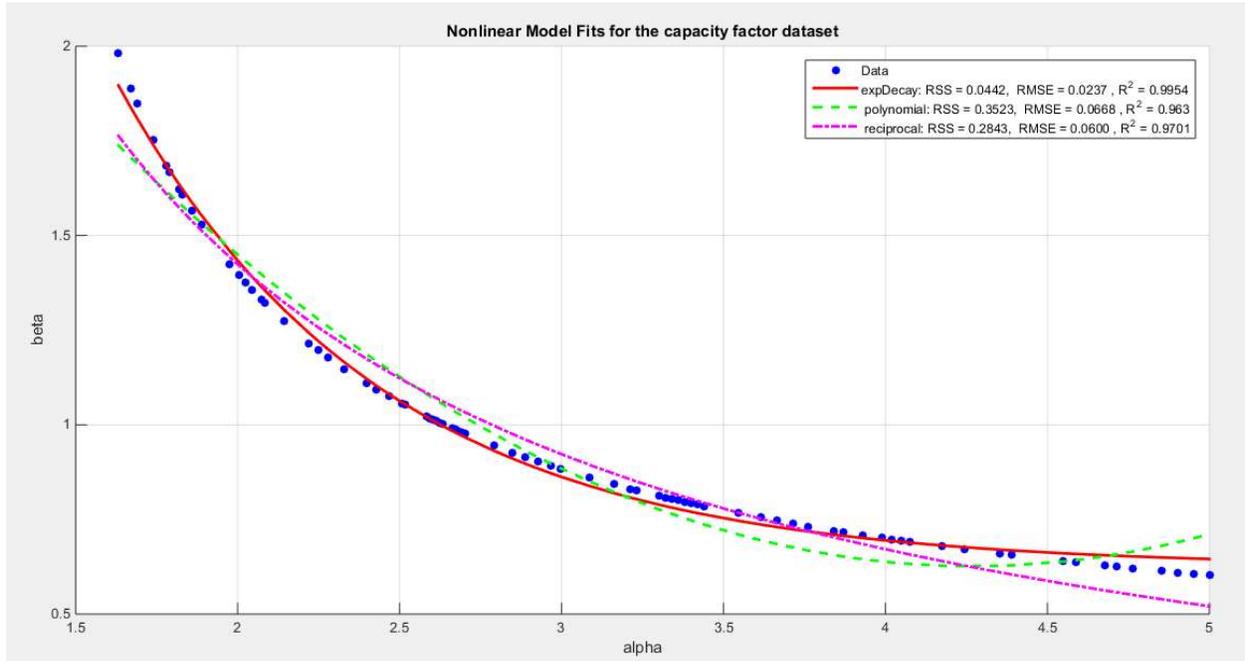

Fig. 12 shows a decreasing convex relationship between alpha and beta. The Nonlinear models for the capacity factor dataset, with residual sum of squares (RSS), R^2 and root of mean square error(RMSE) are illustrated in the figure for each curve, high lightening that the exponential is the best model followed by the reciprocal model then the polynomial ( quadratic).

Table7: the results of the 3 nonlinear models using the capacity data

| metric | Exponential decay | Quadratic-polynomial | Reciprocal |
|---|---|---|---|
| Alpha(CI) | 1.7729(1.7086 , 1.8373 ) | 1.8767(1.8018 , 1.9516) | 1.9217(1.8421 , 2.0012 ) |
| Beta(CI) | 1.6943(1.6098,1.7787) | 1.5412(1.5984 , 1.484 ) | 1.4853(1.4205 , 1.5502 ) |
| V& (SE)a | 0.0248(0.0328) | 0.0336(0.0382) | 0.0379(0.0406) |
| V&(SE)b | 0.0427(0.0431) | 0.0196(0.0292) | 0.0252(0.0331) |
| nLL | -7.6079 | -7.6079 | -7.6079 |
| KS | 0.1518 | 0.1518 | 0.1518 |
| AD | 1.9075 | 1.9075 | 1.9075 |
| CVM | 0.2033 | 0.2033 | 0.2033 |
| AIC | -11.2158 | -11.2158 | -11.2158 |
| CAIC | -10.6158 | -10.6158 | -10.6158 |
| BIC | -8.9448 | -8.9448 | -8.9448 |
| HQIC | 0.5128 | 0.5128 | 0.5128 |
| H₀ = 0 | Fail to reject | Fail to reject | Fail to reject |
| P-value (KS-test) | 0.4074 | 0.4074 | 0.4074 |

The results of the three models are comparable. The variances of parameters are small and are approximately equal between the 3 models. The CI is small. Any model can be chosen.

# Discussion

## Section 4

## Real data analysis using bias-corrected MLE approach

The bias-corrected approach is applied on the first, fifth, and sixth datasets to evaluate the estimated parameters from the variance-corrected approach (Quadratic polynomial model) before and after applying the bias-corrected approach discussed in section 3.

Table 8 shows the values of parameters before and after the correction along with the bias values. Table 9 shows the metrics after correction by equation 32.

Table 8: the bias and values of parameters before and after correction

| | Dwelling | | |
|---|---|---|---|
| | Before correction | bias | After correction |
| alpha | 2.7268 | -0.707 | 2.7975 |
| Beta | 1.6461 | 0.1273 | 1.5188 |
| | Time between failure | | |
| alpha | 2.105 | -0.000656 | 2.1057 |
| Beta | 1.5624 | 0.0171 | 1.5453 |
| | Factors affecting unit capacity | | |
| Alpha | 1.8767 | 0.000459 | 1.8762 |
| beta | 1.5412 | 0.0046 | 1.5366 |

Table 9: Metrics after applying bias-corrected MLE on the mentioned datasets

| Metrics | Dwellings | | Time between failures | | Factors affecting unit capacity | |
|---|---|---|---|---|---|---|
| nLL | -74.0169 | | -19.9277 | | -7.6076 | |
| AIC | -144.0338 | | -35.8553 | | -11.2153 | |
| CAIC | -143.6588 | | -35.2553 | | -10.6153 | |
| BIC | -14.9231 | | -33.5843 | | -8.9443 | |
| HQIC | -3.5348 | | -1.4131 | | 0.5128 | |
| KS | 0.2107 | | 0.1643 | | 0.1536 | |
| AD | 3.2988 | | 0.7135 | | 1.9104 | |
| CVM | 0.6162 | | 0.1371 | | 0.2051 | |
| P(KS) | 0.0766 | | 0.5072 | | 0.4090 | |
| Inverse of K matrix= Var-cov matrix | 0.0016 | -.00056 | 0.0021 | -.00064 | 0.0027 | -.00056 |
| | -.00056 | .00082 | -.00064 | 0.0032 | -.00056 | 0.007 |

Figures 13-15 show the CDF and QQ plot for the above data before and after correction for the mentioned values of the parameters.

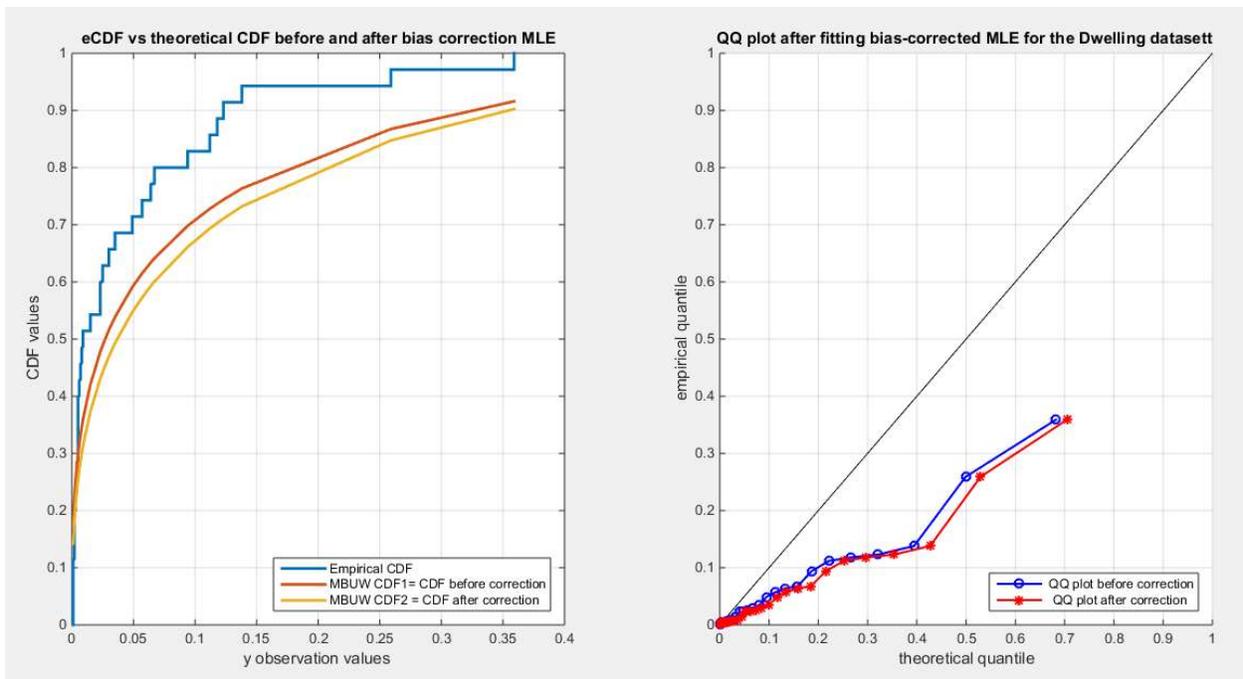

Fig.13 shows the CDF on the right and QQ plot on the left for the Dwelling dataset before and after correction.

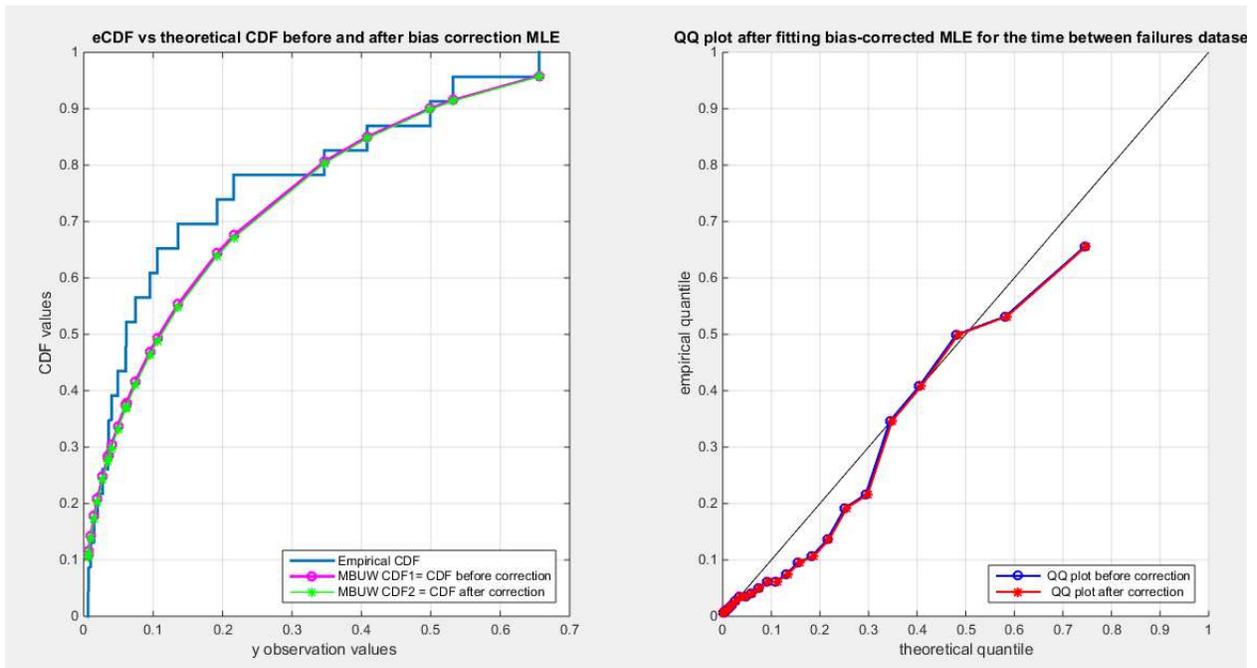

Fig. 14 shows the CDF on the right and QQ plot on the left for the time between failures dataset before and after correction

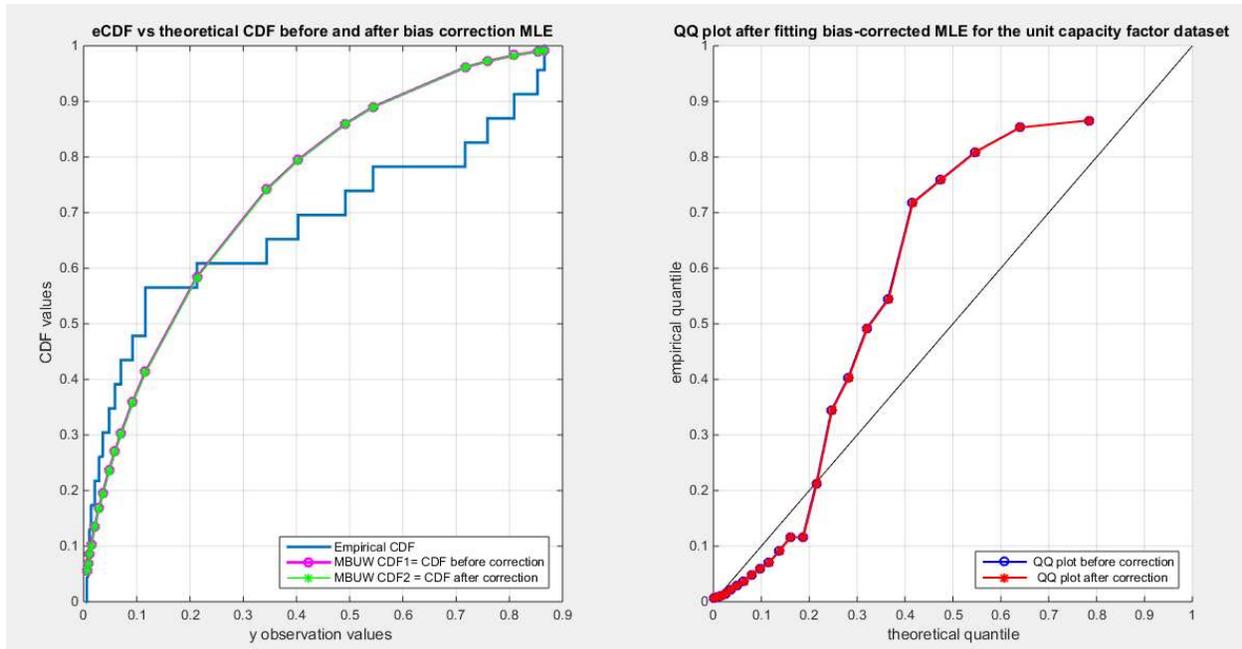

Fig. 15 shows the CDF on the right and QQ plot on the left for the factors affecting unit capacity dataset before and after correction

Table 8 shows minor reduction for the metrics for all the datasets after correction. The variance covariance matrix for each data set exhibit minor variance for each parameter and minimal negative covariance between them reflecting the inverse relationship between them as was illustrated from inspecting the nLL surface in section 3. Figures 13-15 show minor drift between the CDF graph before and after the correction for the dwelling data set and almost no change between the two graphs for the time between failures dataset and factors affecting unit capacity datasets. The inverse of information matrix which represents the variance covariance matrix is positive definite, while the other datasets (quality of support network, voter, and flood datasets have non-positive definite variance covariance matrix, although the KS test results shows fitting of the distribution to these datasets, so the author did not show the results).

The variance-corrected approach leads to marked reduction of the variances and correlation between the parameters. The author used the finite difference approach for all datasets and it resulted into zero correlation between the parameters especially for the data sets with negative correlation and when using the reciprocal models (but the results are unpublished now). Using the delta method induces more reduction than the reduction obtained by the central finite difference method. The resulting uncorrelated estimators with these reduced variances yielded distributions that can fit the different datasets. On the other hand, the bias-corrected approach gives positive definite variance-covariance matrix only in three data sets in contrast to the variance-corrected approach by which all matrices were positive definite. Also the bias-corrected approach heavily depends on the expectations of the higher order derivatives and depends on how to solve for this integration. The method used in solving this integration dramatically impact the information matrix and hence the variance-covariance matrix. The variance-corrected approach yields robust results than methods like generalized method of moments and percentile methods used previously by the author. These latter methods have resulted in large variances and correlated parameters, although their variances are smaller in comparison to the MLE used by Nelder Mead optimizer but still they are inflated. However, using the delta method dramatically reduces the variances and the covariance obtained by GMMs and percentile method. The variance-corrected procedure used in this paper dramatically reduced the variance. If the distribution fits the data set with any estimated parameter values, there is a fixed pattern in the QQ plot. The visualization of the nLL surface for each dataset helps a lot to anticipate the results of the estimation process, the uniqueness, identifiability of the parameters and the correlation between them. The chosen value of the estimated parameters depends on many factors. The most important is that the chosen estimators should have the least variance, goodness of fit test shows good fitness to the data, the metrics like nLL, AIC, CAIC, BIC and HQIC should have the highest negative values. The estimators are preferable to be stable and do not show marked changes to the initial guess especially when using iterative techniques which is the actual case.

# Section 5

## Conclusion

The variance-corrected approach and the bias-corrected approach are valuable techniques used after initial estimation of the MBUW parameters by approaches like MLE using whatever optimizer according to the data characteristics or using generalized methods of moments or percentile methods as these two procedures do not control for correlation between the parameters. So in real world practice, one can start with MLE using Nelder Mead optimization or any other optimizer and if the data characteristics have large impact on the variance, one can shift to methods like generalized method of moments or percentile methods which further reduce the variance but still their variances are large. Visualizing the surface of nLL helps a lot to anticipate the parameter values that can cause a good fit of the distribution to the data and also helps choosing the initial guess of the parameters. With the aid of this inspection, variance-corrected and bias-corrected approach ameliorate the inflated variance and help reducing the correlation between the parameters to yield more robust results.

# Future works

For the three datasets exhibiting non positive definite variance-covariance matrix and similar situations, use other numerical integration techniques or Bayesian inference may help in the estimation process.

**Declarations:**
**Ethics approval and consent to participate**
Not applicable.
**Consent for publication**
Not applicable
**Availability of data and material**
Not applicable. Data sharing does not apply to this article as no datasets were generated or analyzed during the current study.
**Competing interests**
The author declares no competing interests of any type.
**Funding**
No funding resources. No funding roles in the design of the study and collection, analysis, and interpretation of data and in writing the manuscript are declared.
**Authors' contribution**
AI carried the conceptualization by formulating the goals, and aims of the research article, formal analysis by applying the statistical, mathematical, and computational techniques to synthesize and analyze the hypothetical data, carried the methodology by creating the model, software programming and implementation, supervision, writing, drafting, editing, preparation, and creation of the presenting work.
**Acknowledgment**
Not applicable

# References

Bartlett, M. S. (1953a). Approximate Confidence Intervals. *Biometrika*, *40*(1/2), 12. https://doi.org/10.2307/2333091

Bartlett, M. S. (1953b). Approximate Confidence Intervals.II. More than one Unknown Parameter. *Biometrika*, *40*(3/4), 306. https://doi.org/10.2307/2333349

Cordeiro, G. M., Da Rocha, E. C., Da Rocha, J. G. C., & Cribari-Neto, F. (1997). Bias-corrected maximum likelihood estimation for the beta distribution. *Journal of Statistical Computation and Simulation*, *58*(1), 21–35. https://doi.org/10.1080/00949659708811820

Cordeiro, G. M., & Klein, R. (1994). Bias correction in ARMA models. *Statistics & Probability Letters*, *19*(3), 169–176. https://doi.org/10.1016/0167-7152(94)90100-7

Cox, D. R., & Snell, E. J. (1968). A General Definition of Residuals. *Journal of the Royal Statistical Society Series B: Statistical Methodology*, *30*(2), 248–265. https://doi.org/10.1111/j.2517-6161.1968.tb00724.x

Cribari-Neto, F., & Vasconcellos, K. L. P. (2002). Nearly Unbiased Maximum Likelihood Estimation for the Beta Distribution. *Journal of Statistical Computation and Simulation*, *72*(2), 107–118. https://doi.org/10.1080/00949650212144

Efron, B. (1982). *The Jackknife, the Bootstrap andother resampling plans* (1–Vol. 38. Philadelphia, PA, USA: SIAM.). SIAM.

Firth, D. (1993). Bias reduction of maximum likelihood estimates. *Biometrika*, *80*(1), 27–38. https://doi.org/10.1093/biomet/80.1.27

Giles, D. E. (2012). A note on improved estimation for the Topp–Leone distribution. *Tech. Rep., Department of Economics, University of Victoria, Econometrics Working Papers.*

Giles, D. E. (2012). Bias Reduction for the Maximum Likelihood Estimators of the Parameters in the Half-Logistic Distribution. *Communications in Statistics - Theory and Methods*, *41*(2), 212–222. https://doi.org/10.1080/03610926.2010.521278

Giles, D. E., Feng, H., & Godwin, R. T. (2013). On the Bias of the Maximum Likelihood Estimator for the Two-Parameter Lomax Distribution. *Communications in Statistics - Theory and Methods*, *42*(11), 1934–1950. https://doi.org/10.1080/03610926.2011.600506

Giles, D. E & H. Feng. (2009). Bias of the maximum likelihood estimators of the two-parameter Gamma distribution revisited. *Tech. Rep., Department of Economics, University of Victoria, Econometrics Working Papers.*

Haldane, J. B. S., & Smith, S. M. (1956). The Sampling Distribution of a Maximum-Likelihood Estimate. *Biometrika*, *43*(1/2), 96. https://doi.org/10.2307/2333582

Haldane, J.B.S. (1953). The estimation of two parameters from a sample. *Sankhyā*, *12*, 313-320.

Iman M.Attia. (2024). Median Based Unit Weibull (MBUW): ANew Unit Distribution Properties. *25 October 2024*, *preprint article, Preprints.org*(preprint article, Preprints.org). https://doi.org/10.20944/preprints202410.1985.v1, also on arXiv:2410.19019

Lagos-Àlvarez, B., M. D. Jiménez-Gamero, & V. Alba-Fernández. (2011). Bias correction in the Type I Generalized Logistic distribution. *Communications in Statistics - Simulation and Computation*, *40*((4)), 511–531.

Lemonte, A. J. (2011). Improved point estimation for the Kumaraswamy distribution. *Journal of Statistical Computation and Simulation*, *81*(12), 1971–1982. https://doi.org/10.1080/00949655.2010.511621

Lemonte, A. J., Cribari-Neto, F., & Vasconcellos, K. L. P. (2007). Improved statistical inference for the two-parameter Birnbaum–Saunders distribution. *Computational Statistics & Data Analysis*, *51*(9), 4656–4681. https://doi.org/10.1016/j.csda.2006.08.016

Ling, X., & Giles, D. E. (2014). Bias Reduction for the Maximum Likelihood Estimator of the Parameters of the Generalized Rayleigh Family of Distributions. *Communications in Statistics - Theory and Methods*, *43*(8), 1778–1792. https://doi.org/10.1080/03610926.2012.675114

Mazucheli, J., & Dey, S. (2018). Bias-corrected maximum likelihood estimation of the parameters of the generalized half-normal distribution. *Journal of Statistical Computation and Simulation*, *88*(6), 1027–1038. https://doi.org/10.1080/00949655.2017.1413649

Millar, R. (2011). *Maximum likelihood estimation and inference.* Chichester, West Sussex, United Kingdom: John Wiley & Sons, Ltd.

Pawitan, Y. (2001). *In all likelihood: Statistical modelling and inference using likelihood*. Oxford: Oxford University Press.

Reath, J. (2016). *Improved parameter estimation of the log-logistic distribution with applications* [PhD. thesis]. Michigan Technological University.

Saha, K., & Paul, S. (2005). Bias-Corrected Maximum Likelihood Estimator of the Negative Binomial Dispersion Parameter. *Biometrics*, *61*(1), 179–185. https://doi.org/10.1111/j.0006-341X.2005.030833.x

Schwartz, J., & Giles, D. E. (2016). Bias-reduced maximum likelihood estimation of the zero-inflated Poisson distribution. *Communications in Statistics - Theory and Methods*, *45*(2), 465–478. https://doi.org/10.1080/03610926.2013.824590

Schwartz, J., Godwin, R. T., & Giles, D. E. (2013). Improved maximum-likelihood estimation of the shape parameter in the Nakagami distribution. *Journal of Statistical Computation and Simulation*, *83*(3), 434–445. https://doi.org/10.1080/00949655.2011.615316

Shenton, L. R., & Bowman, K. (1963). Higher Moments of a Maximum-Likelihood Estimate. *Journal of the Royal Statistical Society Series B: Statistical Methodology*, *25*(2), 305–317. https://doi.org/10.1111/j.2517-6161.1963.tb00511.x

Singh, A. K., Singh, A., & Murphy, D. J. (2015). On Bias Corrected Estimators of the Two Parameter Gamma Distribution. *2015 12th International Conference on Information Technology - New Generations*, 127–132. https://doi.org/10.1109/ITNG.2015.151

Teimouri, M., & Nadarajah, S. (2013). Bias corrected MLEs for the Weibull distribution based on records. *Statistical Methodology*, *13*, 12–24. https://doi.org/10.1016/j.stamet.2013.01.001

Teimouri, M., & Nadarajah, S. (2016). Bias corrected MLEs under progressive type-II censoring scheme. *Journal of Statistical Computation and Simulation*, *86*(14), 2714–2726. https://doi.org/10.1080/00949655.2015.1123709

Wang, M., & Wang, W. (2017). Bias-Corrected maximum likelihood estimation of the parameters of the weighted Lindley distribution. *Communications in Statistics - Simulation and Computation*, *46*(1), 530–545. https://doi.org/10.1080/03610918.2014.970696

Zhang, G., & Liu, R. (2017). Bias-corrected estimators of scalar skew normal. *Communications in Statistics - Simulation and Computation*, *46*(2), 831–839. https://doi.org/10.1080/03610918.2014.980512

Appendix:

The following are the values of alpha and beta obtained from visualizing or inspecting the nLL surface of the quality of support network dataset:

alpha=[0.119639279, 0.129458918, 0.158917836, 0.168737475, 0.178557114, 0.188376754, 0.198196393, 0.227655311, 0.23747495, 0.257114228, 0.266933868, 0.286573146, 0.296392786, 0.306212425, 0.316032064, 0.345490982]

beta = [0.964328657, 1.00240481, 1.112825651, 1.150901804, 1.188977956, 1.227054108, 1.265130261, 1.383166333, 1.4250501, 1.508817635, 1.550701403, 1.638276553, 1.683967936, 1.729659319, 1.779158317, 1.927655311]